\documentclass[prodmode,acmcsur]{acmsmall} % Aptara syntax
\graphicspath{{figures/}}
\usepackage[ruled]{algorithm2e}
\usepackage{caption}
\usepackage{float}
\usepackage{courier}
\usepackage{listings}
\usepackage{afterpage}
\usepackage{epstopdf}
\usepackage{booktabs}
\usepackage{multirow}
\usepackage{color}
\usepackage{pifont}
\usepackage{glossary}
\usepackage{nomencl}
\usepackage[usenames,dvipsnames,svgnames,table,xcdraw]{xcolor}
\usepackage{parcolumns}

\usepackage{makecell}
\usepackage{lettrine}
\usepackage{amsmath}

\usepackage{soul} 
\usepackage{subfig}
\usepackage{indentfirst}
\usepackage{url}
\usepackage{todonotes}
\usepackage{cite}
\usepackage{listings}
\usepackage{graphicx}
\usepackage{wrapfig}
\usepackage{lscape}
\usepackage{rotating}

\usepackage{pdflscape}
\usepackage{afterpage}
\usepackage{fixltx2e}
\usepackage{booktabs}
\usepackage{float}

\renewcommand{\algorithmcfname}{ALGORITHM}
\SetAlFnt{\small}
\SetAlCapFnt{\small}
\SetAlCapNameFnt{\small}
\SetAlCapHSkip{0pt}
\IncMargin{-\parindent}

\acmVolume{0}
\acmNumber{0}
\acmArticle{00}
\acmYear{2017}
\acmMonth{0}

\begin{document}

% Page heads
\markboth{P. Faruki et al.}{Android Code Protection via Obfuscation Techniques:\\
Past, Present and Future Directions}

% Title portion
\title{Android Code Protection via Obfuscation Techniques:\\
Past, Present and Future Directions}

\author{
Parvez Faruki
\affil{Malaviya National Institute of Technology Jaipur, India}
Hossein Fereidooni
\affil{University of Padua, Italy}
Vijay Laxmi
\affil{Malaviya National Institute of Technology Jaipur, India}
Mauro Conti
\affil{University of Padua, Italy}
Manoj Gaur
\affil{Malaviya National Institute of Technology Jaipur, India}
}

\def\arraystretch{1.4}

\begin{abstract}
%\justify
Mobile devices have become ubiquitous due to centralization of private user information, contacts, messages and multiple sensors. Google Android, an open-source mobile Operating System (OS), is currently the market leader. Android popularity has motivated the malware authors to employ set of cyber attacks leveraging code obfuscation techniques. Obfuscation is an action that modifies an application (app) code, preserving the original semantics and functionality to evade anti-malware. Code obfuscation is a contentious issue. Theoretical code analysis techniques indicate that, attaining a verifiable and secure obfuscation is impossible. However, obfuscation tools and techniques are popular both among malware developers (to evade anti-malware) and commercial software developers (protect intellectual rights). We conducted a survey to uncover answers to concrete and relevant questions concerning Android code obfuscation and protection techniques. The purpose of this paper is to review code obfuscation and code protection practices, and evaluate efficacy of existing code de-obfuscation tools. In particular, we discuss Android code obfuscation methods, custom app protection techniques, and various de-obfuscation methods. Furthermore, we review and analyze the obfuscation techniques used by malware authors to evade analysis efforts. We believe that, there is a need to investigate efficiency of the defense techniques used for code protection. This survey would be beneficial to the researchers and practitioners, to understand obfuscation and de-obfuscation techniques to propose novel solutions on Android. 

\end{abstract}


\acmformat{Parvez Faruki, Hossein Fereidooni, Vijay Laxmi, Mauro Conti and Manoj Gaur.}

% At a minimum you need to supply the author names, year and a title.
% IMPORTANT:
% Full first names whenever they are known, surname last, followed by a period.
% In the case of two authors, 'and' is placed between them.
% In the case of three or more authors, the serial comma is used, that is, all author names
% except the last one but including the penultimate author's name are followed by a comma,
% and then 'and' is placed before the final author's name.
% If only first and middle initials are known, then each initial
% is followed by a period and they are separated by a space.
% The remaining information (journal title, volume, article number, date, etc.) is 'auto-generated'.

\begin{bottomstuff}
Author's addresses: P. Faruki {and} V. Laxmi {and} M. Gaur, Computer Engineering Department, Malaviya National Institute of Technology, India; H. Fereidooni {and} M. Conti, Department of Mathematics, University of Padua.
\end{bottomstuff}

\maketitle

\section{INTRODUCTION}
\lettrine[lines=2]{A}{ndroid}, the most popular mobile device OS is currently the market leader~\cite{guardian:2014}~\cite{gartner:2014}. The availability of Internet, Global Positioning System (GPS) and custom apps have increased the popularity of the mobile devices. The official Android market, Google Play is the dominant app distribution platform accessible to all Android devices~\cite{GooglePlay:2012}. Google Play also allows installation of third-party developers and app stores~\cite{Slideme}. The elevated Android popularity has attracted the attention of malware authors employing advanced code obfuscation and protection techniques. The malware authors are propagating encrypted and obfuscated premium-rate SMS malware, evading Google Play security~\cite{McAfee:2012}. On the other hand, app developers are concerned about code misuse; hence they employ code obfuscation, encryption and custom protection techniques. The weak code protection techniques lower the code integrity and escalate the risk of plagiarism and malware attacks. For instance, a rooted device facilitates identification of app internals and evades device security.

\indent In software engineering lexicon, reverse engineering is defined as a set of methods for obtaining the source code from \texttt{APK} archive. The code obfuscation is employed by malware authors to evade anti-malware. In particular, the architecture-neutral compiled Java code is amenable to reverse engineering. The app developers are concerned about safety and protection of the developed intellectual algorithms and data. Malware authors use obfuscation, code encryption, dynamic code loading, and native code execution evading the Google Play protection~\cite{harvester:2015}.

\indent Code obfuscation is reported as a reasonable and easy alternative  compared to the other protection techniques~\cite{Franz:2010:EUP:1900546.1900550}~\cite{TUD-CS-2012-0135}. Code obfuscation is a set of purposeful techniques to render the code unreadable. Code obfuscation transforms the code by changing its physical appearance, preserving the intended program logic and behavior. Furthermore, code obfuscation is useful to protect the software from reverse engineering. Hence, malware developers have already leveraged the obfuscation by developing recent malware apps evading the Play stores and commercial anti-malware~\cite{dendroid,obad}. Android permits app distribution from third-party developers and other third party app stores. Application developers employ obfuscators to protect the proprietary logic and sensitive algorithms to avoid the misuse. The code obfuscation techniques can be used to: (i) protect the intellectual property; (ii) prevent piracy; and (iii) prevent app misuse. The obfuscation techniques employed by malware authors evade the existing commercial anti-malware solutions. The Android development environment has an in-built obfuscator Proguard~\cite{proguard} for app code protection.

\indent A De-obfuscator is required when app source code is not available. De-obfuscation can be used to verify correct execution of obfuscated app. A popular app may disguise as a \texttt{Trojan} with hidden malicious payload. Zhou et al.~\cite{zhou:dissecting} studied 49 Android malware families and reported more than 86\% repackaged malware from 1260 \texttt{APK} files. However, manual analysis is difficult for evaluating 1,490,272 Google Play apps reported in September 2015~\cite{Googleplayapps}. There exist several legal and technical methods to protect the intellectual property of software producers such as encryption, server-side execution, trusted native code, program encoding, and code obfuscation. 

Android is the dominant mobile platform among users and developers~\cite{guardian:2014}. The open nature of Android allows apps from third-party developers and other third-party markets. The popularity is an opportunity for plagiarists to clone, obfuscate and hijack the popular apps with their trojanized versions. Code obfuscation and protection techniques can protect the intellectual property. However, they can also be used to protect malicious code. 

The remainder of this paper is organized as follows. In Section~\ref{sec:the_android_architecture}, we discuss basics of Android platform, execution mechanism, and its difference with Java development model. Section~\ref{sec:purpose} covers types and purpose of code obfuscation with an in-depth description of app protection techniques. Section~\ref{sec_formalwareauthors} elaborates obfuscation and optimization tools used by malware authors. In Section~\ref{sec_protectorspackers}, we discuss the code protection techniques used by code packers. In Section~\ref{sec_bytecodetools}, we explore the reverse engineering tools used for de-obfuscating the protected code. Section~\ref{sec_frd} elaborates the recent trends in obfuscation and possible future research directions. Finally, in Section~\ref{sec:con}, we conclude this survey.
%Section~\ref{section_futureofobfs} discusses stealth obfuscation techniques used by malware authors.

\section{ANDROID OVERVIEW} \label{sec:the_android_architecture}

The Android OS is open-source developed by Google and supported by the Open Handset Alliance (OHA)~\cite{Russello:2013,Faruki:2013}. In the following, we discuss Android OS architecture and app compilation process.

\subsection{Android Architecture}
\begin{figure}[H]
	\centering
	\includegraphics[width=0.9\textwidth]{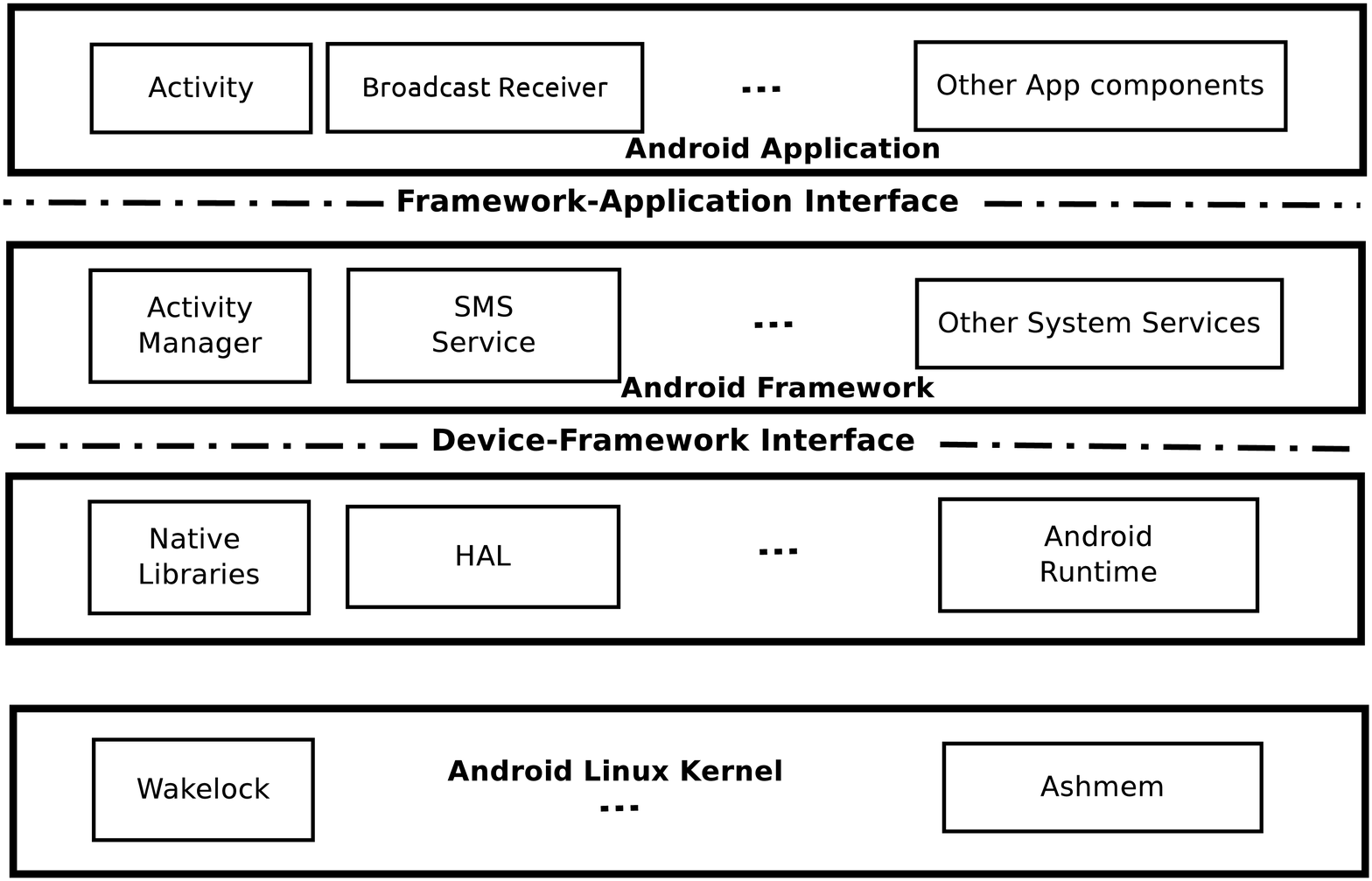}
	\caption{Android Architecture.}
	\label{fig:Android-Architecture}
\end{figure}

\indent The Android software stack has four layers as illustrated in Figure~\ref{fig:Android-Architecture}: (i) linux kernel; (ii) native user-space; (iii) application framework; and (iv) application layer. The base of Android is Linux kernel adapted for limited processing capability, restricted memory and constrained battery availability. The Android platform customized ``vanilla'' kernel for resource constrained mobile devices. The Binder driver for inter-process communication (IPC), Android shared memory (ashmem), and wakelocks are important modifications to suit the Android devices.  

%%%%%%
%%%%%%
\indent Native Space and Application Framework layers form the Android middleware. The bottom layer blocks are the components developed in C/C++. However, the top two layers are implemented in Java. Native components are directly executed on the processor, bypassing the DVM. Hardware Abstraction Layer (HAL) is blurred between the Linux Kernel and Native user-space. Application Framework provides interaction to the developer. Java classes are compiled into Dalvik executable \texttt{.dex} and interpreted by the VM. User apps are at the top most layer. 

\begin{figure}[h!]
	\centering
	\includegraphics[width=0.85\textwidth]{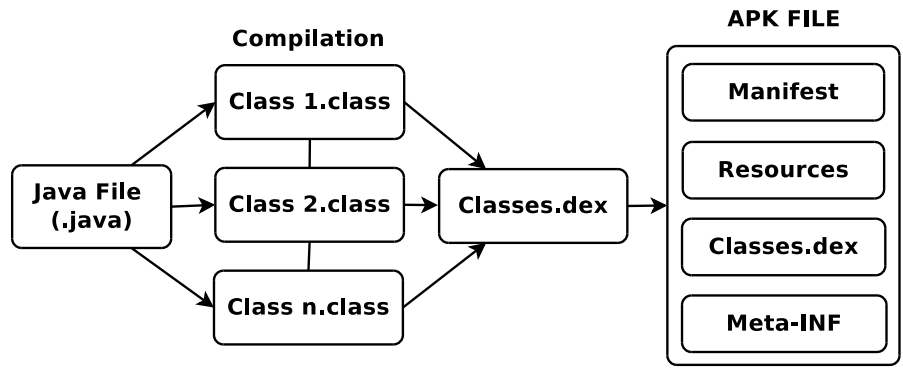}
	\caption{Compilation of Java code to Android \texttt{APK}.}
	\label{fig:android-compilation}
\end{figure}

\indent There are many known open-source and commercial tools to reverse engineer the Android apps. Thus, unprotected apps may unknowingly give away the source code to the attackers for misuse. Android runs the APK files on Dalvik Virtual Machine (DVM), a register-based virtual machine to suit the mobile devices~\cite{Faruki:2013}. The JVM is a stack-based, whereas, DVM is register-based~\cite{Shi:2008:VMS:1328195.1328197}. JVM employs Last In First Out (LIFO) stack with \texttt{PUSH} and \texttt{POP} operations. The DVM stores register-based operands in the CPU registers and requires explicit addressing. Figure~\ref{fig:android-compilation} illustrates the procedure of converting \texttt{Java} source code to an \texttt{APK} archive.

\subsection{Android Compilation}
\indent \texttt{dx} is Android SDK tool that converts Java source code to Dalvik bytecode~\cite{dxtool}. It merges multiple class files into a single \texttt{.dex} file. Android manifest stores name and version of the app, libraries, declared permissions, assets, and other uncompiled resources. The content is merged into a single archive, an Android application PacKage (APK)~\cite{Faruki:2013}. Many open-source and proprietary tools are available for reverse engineering the application. The unprotected apps may unknowingly give away their source code to the attacker, permitting visibility to the internals of the APK. The easy availability of source may lead to loss of revenue, reputation issues, access to intellectual property, and legal liabilities. \newline \newline

\indent The Android KitKat 4.4 introduced Android RunTime (ART) to replace the DVM. The new runtime is proposed to improve the Android OS performance. We briefly compare the Dalvik and ART to underline their importance.

\begin{enumerate}

\item \emph{Dalvik:} \newline
Dalvik VM is the core of Android Dalvik bytecode execution. The Dalvik runtime is based on the Just-In-Time (JIT) compilation that remains independent of the machine code. When a user runs an Android app, the \texttt{.dex} code is compiled to the machine code. Dalvik VM performs JIT compilation and optimization during the app runtime to improve performance. However, the presence of JIT adds latency and memory pressure. Though mobile devices are improving their resources, the new runtime is more efficient in comparison to Dalvik VM.

\item \emph{Android RunTime (ART):} \newline
The Android KitKat version 4.4 introduced an optional runtime Android RunTime (ART) to experimentally replace the DVM to improve performance. In ART, the \texttt{APK} the bytecode is converted to machine code at install time. Ahead-of-time compilation (AOT), a pre-compilation technique, saves the machine code in persistent storage. It loads the machine code at runtime, saving the CPU and memory as compared to DVM. In particular, the \texttt{.dex} file is compiled as \texttt{.oat} file in ELF format. The ART reads the \texttt{.dex} file using \texttt{.dexFile}, \texttt{openDexFileNative} from \texttt{libart.so} library. If the \texttt{oat} file is not found, \texttt{ART} invokes \texttt{dex2oat} tool for compiling \texttt{.dex} to \texttt{.oat}. Otherwise, ART loads the \texttt{.oat} file in memory cache map. Once the \texttt{.oat} file is loaded, ART creates \texttt{OatFile} data structure to store the information. The \texttt{ART} reduces app startup time as code is converted to native at install time, which improves battery life. However, the installation takes more time and space.
\end{enumerate}

\section{CODE OBFUSCATION TECHNIQUES}\label{sec:type-purpose}

\label{sec:purpose}
\begin{figure}[H]
	\centering
	\includegraphics[width=1\textwidth]{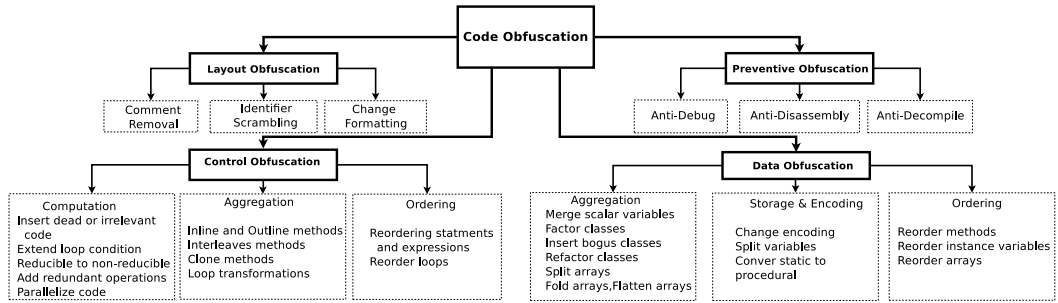}
	\caption{ Obfuscation Classification.}
	\label{fig:Obfuscation-Classification}
\end{figure}

\indent Code obfuscation or mutation techniques alter the code appearance in the existing binary from one generation to another, to evade the anti-malware. Malware authors employ obfuscation techniques to protect malicious logic to evade the anti-malware~\cite{Preda}. The app developers also use obfuscation and code protection methods to protect the code against reverse engineering. Plagiarist and malware authors employ obfuscation to evade the security tools. Malware authors also employ obfuscation to plagiarize the popular and paid apps. Obfuscation provides significant protection and obscures explicit details. Some obfuscation techniques operate directly on the source code; some obfuscate the bytecode. Collberg et al. classify the obfuscation techniques as: (i) Control-flow; (ii) Data; (iii) Layout; and (iv) Preventive obfuscation~\cite{Collbergtaxonomy,conf/isw/Oorschot03}. Figure \ref{fig:Obfuscation-Classification} illustrates detailed outline of existing code obfuscation and protection techniques discussed subsequently. An obfuscator protects the proprietary software and prevents its reverse engineering. The obfuscated program maintains the semantics of the original app. The original and obfuscated version produce the same output when executed. The malware authors use code protection and obfuscation techniques to protect malware logic, algorithms and hide the suspicious information such as strings, domain names and server address to delay malware being detected.

%\begin{sidewaysfigure}
%	\centering
%	\includegraphics[width=1\textwidth]{figures/Obfuscation-Classification.eps}
%	\caption{ Obfuscation Classification.}
%	\label{fig:Obfuscation-Classification}
%\end{sidewaysfigure}

\indent A de-obfuscator restores the original app from the obfuscated code with reverse engineering tools. The reconstruction of the source may not be possible. For example, identifier renaming obfuscated variables cannot be restored to their original names once changed by the obfuscator. According to~\cite{Low98javacontrol}, code obfuscation or transformation is the most suitable technical protection for type-safe language like Java. According to Udupa et al.~\cite{Udupa05deobfuscation:reverse}, surface obfuscation affects the syntax of the target program. However, it is not possible to hide the code semantics. If an identifier is renamed, the program output remains the same. The deep transformation obfuscates the control flow of target program~\cite{325schulz}.

\indent \emph{\textbf{Definition:}} Collberg et al. define the obfuscating transformation as the conversion $\tau$ of a source app $A$ into the target app $A'$~\cite{Collbergtaxonomy} :  
\begin{align*}
 A\stackrel{\mathrm{\tau}}{\longrightarrow}A^{\prime}. 
\end{align*}  
Transformed app $A'$ is an obfuscated version of the original program $A$ if:
\begin{itemize} 
\item App $A$ does not terminate, or ends with an error condition, then $A'$ may or may not terminate~\cite{Collbergtaxonomy}.
\item Otherwise, $A'$ must terminate and generate the output similar to $A$~\cite{uspatent}.
\end{itemize}

\indent Android apps are distributed as archive APK files available from: (i) Google Play; (ii) Third-party app marketplaces; and (iii) Android debug bridge. Once installed, the apps from the devices can be also accessed with Android Developer Tools (ADT). These APK files can be reverse engineered to the Dalvik bytecode from \texttt{classes.dex}, the app executable file. In the following subsection, existing code obfuscation approaches are detailed according to the outline made in Figure~\ref{fig:Obfuscation-Classification}. \newline \newline \newline

\subsection{CONTROL FLOW OBFUSCATION}
\indent Control flow obfuscation aims to confuse the analyst by breaking up the control flow of the source code. Functional blocks that belong together are broken apart, and functional blocks that do not belong together are intermingled to confuse the reverse engineering. The control flow transformation changes the execution paths of a program, still maintains the original functionality. The control flow obfuscation is categorized as Aggregation, Computation and Ordering obfuscation techniques. The techniques can be further classified into following sub-categories as illustrated in Figure~\ref{fig:Obfuscation-Classification}.

\subsubsection{\textbf{Control flow computation }} The control flow obfuscation techniques hide the control flow and append additional code and complicate \texttt{APK} disassembly. The code insertion can be an additional method or irrelevant code~\cite{EgilAspervick}. However, the recent compilers remove unused code for execution efficiency during the code optimization phase. To counter this, one can insert irrelevant bytecode such as \texttt{PUSH} or \texttt{POP} within the high-level code. Hence, it is not removed. Computation control obfuscation can be broken down as described in the following:

\begin{enumerate}
\item {\emph{Inserting Dead or Irrelevant Code:}} The dead code block can never be reached; hence, is never executed. Inserting dead-code statements increases size of code and analysis time. A programmer can insert code block that is never executed~\cite{conf/iait/Kovacheva13}. For example, one can include extra methods or irrelevant statement blocks~\cite{StampWong06_HuntingForMetamorphicEngines}. For instance, the code snippet before the Add Dead-code Switch Statements (ADSS)~\cite{Faruki:2013} obfuscation is illustrated in listing~\ref{before_adss}. The Java bytecode switch construct can be used to insert control switch that is never executed~\cite{conf/cc/BatchelderH07}. However, the switch increases the connectedness and complexity of the method. Thus, the obfuscation evades the decompiler that cannot remove the dead switch. Listing~\ref{after_adss} illustrates the ADSS obfuscation~\cite{conf/cc/BatchelderH07}. \newline \newline \newline 

\noindent\begin{minipage}{.45\textwidth}
\definecolor{mygreen}{rgb}{0,0.6,0}
\lstset{
    captionpos=b,
    caption={before ADSS obfuscation},
    label=before_adss,
    keywordstyle=\color{blue},
    commentstyle=\color{mygreen},
    basicstyle=\fontsize{7}{8}\selectfont\ttfamily,
    frame=single,
    numbers=left,
    belowcaptionskip=2\baselineskip,
    tabsize=2,
}
\begin{lstlisting}[language=Java]
//before ADSS obfuscation
if (writeImage != null) { 
	try { 
	
		File file = new File("out"); 
		ImageIO.write(writeImage, "png", file);
	} 
	catch (Exception e) { 
	System.exit(1); 
	} 
} 
 System.exit(0);










\end{lstlisting}

\end{minipage}\hfill
\begin{minipage}{.45\textwidth}
\definecolor{mygreen}{rgb}{0,0.6,0}
\lstset{
    captionpos=b,
    caption={ADSS obfuscated code~\protect\cite{soap:2014}},
    label=after_adss,
    keywordstyle=\color{blue},
    commentstyle=\color{mygreen},
    basicstyle=\fontsize{7}{8}\selectfont\ttfamily,
    frame=single,
    belowcaptionskip=2\baselineskip,
    tabsize=2,
}
\begin{lstlisting}[language=Java]
// ADSS obfuscated code
if(obj != null) {
 try { 
  ImageIO.write((RenderedImage)obj,png,
new File(out));
    }
  catch(Exception exception2)  {
   i += 2;
   System.exit(1); 
    }
}
label_167:
{ while(lI1.booleanValue() == ___) 
  { switch (i) { 
    default: break;
    case 3: break label_167;
    System.exit(1); 
    continue;
       }
    }
    System.exit(0);
}
\end{lstlisting}
\end{minipage}
\vspace{-0.1mm}
%\vspace{-0.2cm}

\item \emph{Extend Loop condition:} Rewriting the test condition as a complex loop function introduces obfuscation in the code. It can be accomplished by extending the loop condition with the addition of more test cases having no effect on the result. The code in {Listing~\ref{after_loopextend}} illustrates the extension of a simple \texttt{if condition}. \newline \newline

\begin{minipage}{.45\textwidth}
\definecolor{mygreen}{rgb}{0,0.6,0}
\lstset{
    captionpos=b,
    caption={code before Loop extension},
    label=before_loopextend,
    keywordstyle=\color{blue},
    commentstyle=\color{mygreen},
    basicstyle=\fontsize{7}{8}\selectfont\ttfamily,
    frame=single,
    numbers=left,
    belowcaptionskip=2\baselineskip,
    tabsize=2,
}
\begin{lstlisting}[language=Java]
// Before loop extension

int x = 1;

if (x > 200)
{
	...
	x ++;
	Foo(x)
}

\end{lstlisting}

\end{minipage}\hfill
\begin{minipage}{.45\textwidth}
\definecolor{mygreen}{rgb}{0,0.6,0}
\lstset{
    captionpos=b,
    caption={Loop extension obfuscation},
    label=after_loopextend,
    keywordstyle=\color{blue},
    commentstyle=\color{mygreen},
    basicstyle=\fontsize{7}{8}\selectfont\ttfamily,
    frame=single,
    belowcaptionskip=2\baselineskip,
    tabsize=2,
}
\begin{lstlisting}[language=Java]
// After loop extension obfuscation

int x = 1;
while (x> 200 || x%200==0)
{
	...
	x ++;

// calling function
Foo(x)

}
\end{lstlisting}
\end{minipage}
\vspace{-0.05mm}

\item \emph{Reducible to Non-Reducible flow-graph obfuscation (RNR)}~\cite{uspatent}: A reducible flow graph can be made complex by turning it reducible to non-reducible. The Java bytecode has a \texttt{goto} instruction. However, the Java language does not have a corresponding \texttt{goto} statement~\cite{patents}. Hence, a plagiarist can misuse the \texttt{goto} bytecode and obfuscate with an arbitrary control-flow transformation.  Java language can only express a structured control flow. Hence, ``the control flow graphs produced by the Java programs is always reducible. However, Java bytecode can express this as non-reducible flow graphs, thus obfuscating reducible flow graph to non-reducible''~\cite{patents}. According to~\cite{Collbergtaxonomy}, this is achieved by converting a structured loop into a loop with multiple headers. Listing~\ref{before_reducible} and~\ref{after_reducible} illustrate the effect of RNR obfuscation. \newline \newline 

\begin{minipage}{.45\textwidth}
\definecolor{mygreen}{rgb}{0,0.6,0}
\lstset{
    captionpos=b,
    caption={before RNR Obfuscation.},
    label=before_reducible,
    keywordstyle=\color{blue},
    commentstyle=\color{mygreen},
    basicstyle=\fontsize{7}{8}\selectfont\ttfamily,
    frame=single,
    numbers=left,
    belowcaptionskip=2\baselineskip,
    tabsize=2,
}
\begin{lstlisting}[language=Java]
// Before:

Statement 1;
while (condition1)
{

	Statement2;
}
\end{lstlisting}

\end{minipage}\hfill
\begin{minipage}{.45\textwidth}
\definecolor{mygreen}{rgb}{0,0.6,0}
\lstset{
    captionpos=b,
    caption={RNR Obfuscated code.},
    label=after_reducible,
    keywordstyle=\color{blue},
    commentstyle=\color{mygreen},
    basicstyle=\fontsize{7}{8}\selectfont\ttfamily,
    frame=single,
    belowcaptionskip=2\baselineskip,
    tabsize=2,
}
\begin{lstlisting}[language=Java]
// After:
Statement 1;
if(condition2) {
	Statement2;
	while(condition1){
		Statement2;
	}
else {
	while(condition1){
		Statement2;
	}
} 
\end{lstlisting}
\end{minipage}
\vspace{-0.5 mm}

\item \emph{Add Redundant Operands:} Appending insignificant terms within the code during basic calculations hinders reverse engineering. For example, let us assume an integer variable $p$ which stores the product of two integer variables $a$ and $b$. The code listings~\ref{before_redundant} and~\ref{after_redundant} illustrates the code snippets before and after redundant operators obfuscation. The transformed code generates exactly the same output. However, the obfuscated snippet appears complex during the analysis.

\begin{minipage}{.45\textwidth}
\definecolor{mygreen}{rgb}{0,0.6,0}
\lstset{
    captionpos=b,
    caption={before redundant operators},
    label=before_redundant,
    keywordstyle=\color{blue},
    commentstyle=\color{mygreen},
    basicstyle=\fontsize{7}{8}\selectfont\ttfamily,
    frame=single,
    numbers=left,
    belowcaptionskip=2\baselineskip,
    tabsize=2,
}
\begin{lstlisting}[language=Java]
//original code 

public int sum{

int a,b = 5, 7;
int p;
p = a * b;
System.out.println(``Product ='' + p); 

}

\end{lstlisting}

\end{minipage}\hfill
\begin{minipage}{.45\textwidth}
\definecolor{mygreen}{rgb}{0,0.6,0}
\lstset{
    captionpos=b,
    caption={redundant operators},
    label=after_redundant,
    keywordstyle=\color{blue},
    commentstyle=\color{mygreen},
    basicstyle=\fontsize{7}{8}\selectfont\ttfamily,
    frame=single,
    belowcaptionskip=2\baselineskip,
    tabsize=2,
}
\begin{lstlisting}[language=Java]
//Redundant Operators Obfuscation

public int sum{
int a,b = 5,7;
double i,j = 0.0005, 0.0007;
double p;
p = (a * b) + (i*j);
System.out.println(``Product =''+ (int) p); 
}
\end{lstlisting}
\end{minipage}

\item \emph{Parallelize Code:} The introduction of threads can affect the readability due to increased code complexity. The parallelization improves the performance. However, the motivation in this case is to hide the correct code flow. Collberg et al.~\cite{Collbergtaxonomy} suggested the following techniques: (i) Creating dummy processes; or (ii) Splitting sequential sections of a program into multiple concurrent and parallel processes.
%\vspace{0.1 cm}

\end{enumerate}

\subsubsection{\textbf{Control-flow Aggregation (CFA) }}The CFA alters the program statements grouping~\cite{plasmanthesis}. CFA can be further classified into:
   \begin{enumerate}

   \item \emph{Inline and Outline methods:} In Java, replacing a method call by its actual body (inlining) make the code complicated and difficult to understand. Code optimizers use these techniques. It also is a useful obfuscation transformation. Code inlining removes procedural abstraction from the program. Conversely, outlining selects a group of statements in a procedure and re-use them to generate a sub-procedure. For instance, inlining two procedures \texttt{A} and \texttt{B} necessitates the calling one after the other and outlining a portion of the combined code inside a new procedure.

   \item \emph{Method Interleaving (MI):} Identifying an interleaved method is a difficult reverse engineering task. MI merges the body and parameter list of different methods. Furthermore, it adds parameter to discriminate between calls to the individual methods~\cite{Collbergtaxonomy,uspatent,patents}.

   \item \emph{Method Cloning (MC):}
A reverse engineer examines the body and signature of a subroutine to determine code functionality, a necessary step for reverse engineering. One can create function clones and further introduce repetitive calls to such functions. 
%\vspace{0.2 cm}
   \item \emph{Loop Transformations (LT):} In~\cite{Collbergtaxonomy}, authors observed that some transformations increase the code complexity. Such obfuscation techniques break down the iteration space to fit the inner loop cache~\cite{uspatent}. The availability of compile-time loop bounds turns the loop to a compound body with several loops known as loop fission~\cite{uspatent}.
   \end{enumerate}

\subsubsection{\textbf{Control-flow ordering (CFO)}} Control flow ordering obfuscation changes the execution order of the code statements. For instance, loops can be iterated backward rather than forward. Control ordering obfuscations can be categorized into:

  \begin{enumerate}

   \item \emph{Reorder Statements and Expressions (RSE):} Changing the order of statements and expressions has a significant effect on the Dalvik bytecode. It can disrupt the link between Java source and corresponding Dalvik bytecode~\cite{conf/iait/Kovacheva13}. 

   \item \emph{Reorder Loop (RL):} RL transformation can be run backward to confuse the analysis. Code Listing~\ref{before_loopreversal} and~\ref{after_loopreversal} illustrate the usage of loop reversal obfuscation.
\vspace{0.2cm}

\begin{minipage}{.45\textwidth}
\definecolor{mygreen}{rgb}{0,0.6,0}
\lstset{
    captionpos=b,
    caption={before loop reversal},
    label=before_loopreversal,
    keywordstyle=\color{blue},
    commentstyle=\color{mygreen},
    basicstyle=\fontsize{7}{8}\selectfont\ttfamily,
    frame=single,
    numbers=left,
    belowcaptionskip=2\baselineskip,
    tabsize=2,
}
\begin{lstlisting}[language=Java]
// Original code
x = 0;
while (x < maxNum){
	i[x] += j[x];
	x++;
}

\end{lstlisting}

\end{minipage}\hfill
\begin{minipage}{.45\textwidth}
\definecolor{mygreen}{rgb}{0,0.6,0}
\lstset{
    captionpos=b,
    caption={loop reversal obfuscation},
    label=after_loopreversal,
    keywordstyle=\color{blue},
    commentstyle=\color{mygreen},
    basicstyle=\fontsize{7}{8}\selectfont\ttfamily,
    frame=single,
    belowcaptionskip=2\baselineskip,
    tabsize=2,
}
\begin{lstlisting}[language=Java]
// Loop reversal to change the control flow
x = maxNum;
while (x > 0)
{
	x--;
	i[x] += j[x];
}
\end{lstlisting}
\end{minipage}
\vspace{0.2 cm}

\vspace{-1cm}

\subsubsection{\textbf{Control-flow flattening (CFF)}} CFF transforms the source code such that static analysis cannot determine the targets of branches. In this technique,  the basic blocks of a program have same predecessor and successor. During execution, the actual control flow is controlled by a \emph{dispatch variable}. A switch block has dispatch variable and a jump table to switch indirectly towards intended successor. At runtime, the value of dispatch variable decides which code block to be executed next.  Figure~\ref{fig:Control Flow Transformations.}(a) illustrates the control-flow graph of Listing~\ref{cff_snippet} and Figure~\ref{fig:Control Flow Transformations.}(b) shows the flattened control flow of the mentioned program. Compilers primarily use such methods during code optimization. However, malware authors leverage this technique to hide the semantic as well as syntax structure of a malware \texttt{APK}. This obfuscation technique prevents or at least delays manual analysis. 
\vspace{0.2cm}

\definecolor{mygreen}{rgb}{0,0.6,0}
\lstset{
    captionpos=b,
    caption={control flow flattening},
    label=cff_snippet,
    keywordstyle=\color{blue},
    commentstyle=\color{mygreen},
    basicstyle=\fontsize{7}{8}\selectfont\ttfamily,
    frame=single,
    numbers=left,
    belowcaptionskip=2\baselineskip,
    tabsize=2,
}
\begin{lstlisting}[language=Java]
public int fill( int a , int b ) 
{ 
	int diff = 0;
	if ( a > b ) { diff = a-b ;
		do 
		{ b ++;
		} while ( a ! = b ) ;
	} 
	else { diff = b-a ;
		do { a ++;
	} while ( a ! = b ) }
	return diff ;
}
\end{lstlisting}
\vspace{-0.05 mm}
\begin{figure}[h!]
	\centering
	\includegraphics[width=0.81\textwidth]{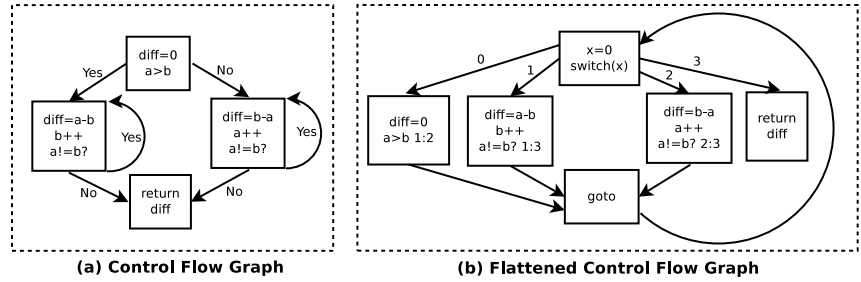}
	\caption{CFG of listing~\ref{cff_snippet}}
	\label{fig:Control Flow Transformations.}
\end{figure} 

\vspace{-1cm}
\subsection{DATA OBFUSCATION}
\indent Data obfuscation techniques can modify the structure of an APK. They can be classified into following sub-categories as illustrated in Figure~\ref{fig:Obfuscation-Classification}.

\subsubsection{\textbf{Data aggregation}} This obfuscation changes the data grouping. In the following, we list some sub-categories of this technique:

   \begin{enumerate}
   \item \emph{Merging Scalar Variables:} This obfuscation involves merging two or more scalar variables into a single variable. For example, $m$ scalar variables \emph{Var\textsubscript{1}, Var\textsubscript{2},Var\textsubscript{3},..,Var\textsubscript{k}} can be merged into a single array variable \emph{Var\textsubscript{m}}. Variables, in a way similar to arrays or integers, can be merged or even promoted as objects~\cite{Collbergtaxonomy,uspatent,patents}. As a simple example, consider merging two 32-bit integer variables X and Y into a 64-bit variable Z. 

   \item \emph{Class Transformations (CT):} Class transformations can be leveraged to make the program analysis difficult. One good way of achieving this is to use inheritance and interfaces to create deep class hierarchies to build a complex distributed application. Further, one or more dummy classes/methods can confuse the reverse engineer.
%\vspace{0.1cm}
   \item \emph{Array Transformations (AT):} Array transformation is an effective obfuscation technique to convert the readable string information as unreadable~\cite{uspatent}. The AT: (i) splits an array; (ii) merge two or more arrays; (iii) flattens an array dimensions;  or (iv) folds or increases  the array dimensions.
    \end{enumerate}
\vspace{-0.3cm}
\subsubsection{\textbf{Data Storage and Encoding (DSE)}} The DSE affects how the data is stored and interpreted. Such methods obscure the data structures within the programs. Data storage obfuscation converts a local variable into a global. Furthermore, data encoding obfuscations replaces an integer variable \texttt{i} with an expression \texttt{x*i+y}. In the following, we discuss such techniques.
%\vspace{0.2cm}  
  \begin{enumerate}
  \item \emph{Change Encoding (CE):} Programmers follow some standard conventions to write code. The encoding transformation techniques exploit this fact. The more transformations we employ, the less likelihood to understand the code. In fact, changing encoding reshapes the data into less natural forms. For example, we can replace all the references initializing an index variable \texttt{i}, with the expression \texttt{i= x*i+y}, where \texttt{x=6} and \texttt{y=5}. When the code needs to use the index value, the obfuscator inserts the expression \texttt{(i-5)/6}. Finally, instead of incrementing the variable by one, add six to the value. The obfuscation scales and offsets the index from the desired value to compute the real index. Listings~\ref{Before-encoding-obfuscation} and~\ref{After-encoding-obfuscation} illustrates encoding. \newline

\vspace{0.3cm}

\begin{minipage}{.4\textwidth}
\definecolor{mygreen}{rgb}{0,0.6,0}
\lstset{
    captionpos=b,
    caption={before encoding obfuscation},
    label=Before-encoding-obfuscation,
    keywordstyle=\color{blue},
    commentstyle=\color{mygreen},
    basicstyle=\fontsize{7}{8}\selectfont\ttfamily,
    frame=single,
    numbers=left,
    belowcaptionskip=2\baselineskip,
    tabsize=2,
}
\begin{lstlisting}[language=Java]
//Before:
int i = 1;
while (i <= 100) 
{
	result = arr[i-1];
	i++;
}
\end{lstlisting}

\end{minipage}\hfill
\begin{minipage}{.4\textwidth}
\definecolor{mygreen}{rgb}{0,0.6,0}
\lstset{
    captionpos=b,
    caption={After encoding obfuscation},
    label=After-encoding-obfuscation,
    keywordstyle=\color{blue},
    commentstyle=\color{mygreen},
    basicstyle=\fontsize{7}{8}\selectfont\ttfamily,
    frame=single,
    belowcaptionskip=2\baselineskip,
    tabsize=2,
}
\begin{lstlisting}[language=Java]
//After:
int i = 11;
while (i <= 605) 
{	result = arr[(i-5)/6];
	i+=6;
}
\end{lstlisting}
\end{minipage}

\vspace{-0.2cm}
   \item \emph{Class Transformations:} Class transformations can be leveraged to make the program analysis difficult. One good way of achieving this is to use inheritance and interfaces to create deep class hierarchies that make it more difficult to understand the application. Inserting a bogus class can confuse the reverse engineer.
\vspace{0.1cm}
   \item \emph{Array Transformations:} There exist many different types of transformations that can be devised to obscure the operations performed on an array. These transformations consist of splitting an array, merging two or more arrays, flattening an array (i.e., decrease the dimensions of the array), and folding the array (i.e., increase the dimensions of the array).
    \end{enumerate}

The data modification techniques can be used to evade the string based malware detectors. A simple example illustrated in Listing~\ref{before_integerobfuscation} and~\ref{after_integerobfuscation} demonstrates data modification obfuscation.

\begin{minipage}{.4\textwidth}
\definecolor{mygreen}{rgb}{0,0.6,0}
\lstset{
    captionpos=b,
    caption={before integer obfuscation},
    label=before_integerobfuscation,
    keywordstyle=\color{blue},
    commentstyle=\color{mygreen},
    basicstyle=\fontsize{7}{8}\selectfont\ttfamily,
    frame=single,
    numbers=left,
    belowcaptionskip=2\baselineskip,
    tabsize=2,
}
\begin{lstlisting}[language=Java]
// original declaration
	int a = 30													.
\end{lstlisting}

\end{minipage}\hfill
\begin{minipage}{.4\textwidth}
\definecolor{mygreen}{rgb}{0,0.6,0}
\lstset{
    captionpos=b,
    caption={after integer obfuscation},
    label=after_integerobfuscation,
    keywordstyle=\color{blue},
    commentstyle=\color{mygreen},
    basicstyle=\fontsize{7}{8}\selectfont\ttfamily,
    frame=single,
    belowcaptionskip=2\baselineskip,
    tabsize=2,
}
\begin{lstlisting}[language=Java]
// obfuscating integer declaration
class bar {
public int getValue () {
return 30; }
}
Foo f = new bar () ;
int a = f.getValue () ;
\end{lstlisting}
\end{minipage}
\vspace{-0.05mm}

  \item \emph{Split Variables (SV):} Boolean variables can be replaced by a boolean expression. The relevant example is illustrated in Table~\ref{table:Lookup-Table}. Variable $b$ in the original code is expressed as$f(b_{1},b_{2})$. In this example, function \texttt{F} is \texttt{XOR}. However, it can be generalized to any function with any number of variables. This also adds another layer of obfuscation due to the fact that some assignments of a value have different results. Let us say $b$ = True. We can further split the variable into $b_{1}$ = 0 and $b_{2}$ = 1 in the lookup Table~\ref{table:Lookup-Table} and convert it to original Boolean value. 
%\vspace{-0.3cm}
\begin{table}[H]
\centering
    \small    
    \begin{tabular}{| l | l | l | }
    \hline
 %  \textbf{F\textsuperscript{-1}(b)} &  & \textbf{F(b\textsubscript{1},b\textsubscript{2})} \\ 
    {\centering \textbf{$b\textsubscript{1}$}} & {\centering \textbf{$b\textsubscript{2}$}} & {\centering \textbf{$b$}}=\textbf{$F(b\textsubscript{1}\oplus b\textsubscript{2}$)}  \\ \hline
    0 & 0 & \emph{False} \\ \hline
    0 &  1 &  \emph{True}\\ \hline
    1 & 0 & \emph{True} \\ \hline
    1 &  1 &  \emph{False}\\ \hline
    \end{tabular}
    \caption{Lookup table to split variables~\protect\cite{palisade}}
    \label{table:Lookup-Table}
\end{table}
\vspace{-0.3cm}
To split the variable $b$ into $b\textsubscript{1}$ and $b\textsubscript{2}$, we have to define: (i) a function, $F(b\textsubscript{1},b\textsubscript{2})$ that maps $b\textsubscript{1}$ and $b\textsubscript{2}$ to variable $b$, (ii) to the inverse function, $F\textsuperscript{-1}(b)$ that maps $b$ to $b\textsubscript{1}$ and $b\textsubscript{2}$, and (iii) new operations defined on $b\textsubscript{1}$ and $b\textsubscript{2}$. $b$ has been split into two shorter integer variables $b\textsubscript{1}$ and $b\textsubscript{2}$. If $b\textsubscript{1}$=$b\textsubscript{2}$=0 or $b\textsubscript{1}$=$b\textsubscript{2}$=1 then $b$ is false, otherwise, $b$ is True. Boolean $b$ are masked as arithmetic operations on the integers $b\textsubscript{1}$ and $b\textsubscript{2}$ with the split variable technique~\cite{palisade}.
%\vspace{0.15cm}
     \item \emph{Convert Static to Procedural Data:} Strings store critical information such as copyright information, license key, and software expiry date. If the static string information is converted to procedural data, reverse engineering becomes complicated. A simple way obfuscate is to convert the string to a program that computes the string~\cite{Wroblewski}.
   \end{enumerate}

\subsubsection{\textbf{Data Ordering}} This obfuscation alters the data ordering. Ordering transformation randomizes data declaration order within a program. ``An array stores a list of integer numbers. The array has the \emph{i-th} element in the list at position \emph{i}. A function \texttt{f(i)} can be used to determine the position of the \emph{i-th} element in the list''~\cite{Preda,Low98javacontrol}. Randomizing the declaration order impedes reverse engineering process~\cite{Low98javacontrol,StampWong06_HuntingForMetamorphicEngines}.
  
  \begin{enumerate}
 % \vspace{0.15cm}
   \item \emph{Reorder Methods:}
randomize the declarations of methods within the code to harden the reverse engineering. 
 %\vspace{0.15cm} 
    \item \emph{Reorder Arrays:} randomize the order of parameters to methods and use a mapping function to reorder data within arrays. 
  %   \vspace{0.15cm}  
    \item \emph{Reorder Instance Variables:} randomize the declarations of instance variables within the class.
   \end{enumerate}

\begin{minipage}{.45\textwidth}
\definecolor{mygreen}{rgb}{0,0.6,0}
\lstset{
    captionpos=b,
    caption={before aggregation},
    label=before_aggregation,
    keywordstyle=\color{blue},
    commentstyle=\color{mygreen},
    basicstyle=\fontsize{7}{8}\selectfont\ttfamily,
    frame=single,
    numbers=left,
    belowcaptionskip=2\baselineskip,
    tabsize=2,
}
\begin{lstlisting}[language=Java]
// original code
int a = 7 + 70/2 + 1 ;
int b = "testing".length();
\end{lstlisting}

\end{minipage}\hfill
\begin{minipage}{.45\textwidth}
\definecolor{mygreen}{rgb}{0,0.6,0}
\lstset{
    captionpos=b,
    caption={aggregation obfuscation},
    label=after_aggregation,
    keywordstyle=\color{blue},
    commentstyle=\color{mygreen},
    basicstyle=\fontsize{7}{8}\selectfont\ttfamily,
    frame=single,
    belowcaptionskip=2\baselineskip,
    tabsize=2,
}
\begin{lstlisting}[language=Java]
// obfuscated code
int a = 43;
int b = 7;
\end{lstlisting}
\end{minipage}
\vspace{0.3cm}

\subsection{LAYOUT OBFUSCATION}
\indent Layout transformation is used to modify the source and binary structure of a program. Different layout transformation techniques are illustrated in Figure~\ref{fig:Obfuscation-Classification}. The Layout Obfuscation can be classified as: (i) identifier scrambling; (ii) output format changes; (iii) comments, or debug information.

Replacing the identifier names with mandarin non-alphabetical characters increases the code complexity. For example, in code Listing~\ref{before_renaming} the SmsManager Class encrypts the string before sending it over the network. However, the original identifiers names are amenable to reverse engineering. To avoid code inference, oBad Trojan employs identifier mangling obfuscation in Listing~\ref{after_renaming}. \newline

\begin{minipage}{.4\textwidth}
\definecolor{mygreen}{rgb}{0,0.6,0}
\lstset{
    captionpos=b,
    caption={before obfuscation},
    label=before_renaming,
    keywordstyle=\color{blue},
    commentstyle=\color{mygreen},
    basicstyle=\fontsize{7}{8}\selectfont\ttfamily,
    frame=single,
    numbers=left,
    belowcaptionskip=2\baselineskip,
    tabsize=2,
}
\begin{lstlisting}[language=Java]
package iAmDriving{
	public class LetsNavigate{...}
}
package iAmConnecting {
	public class MyBluetoothHandle{..}
}
package userIsActive{
	public class MainActivity{...}
}
\end{lstlisting}

\end{minipage}\hfill
\begin{minipage}{.4\textwidth}
\definecolor{mygreen}{rgb}{0,0.6,0}
\lstset{
    captionpos=b,
    caption={identifier renaming~\protect\cite{325schulz}},
    label=after_renaming,
    keywordstyle=\color{blue},
    commentstyle=\color{mygreen},
    basicstyle=\fontsize{7}{8}\selectfont\ttfamily,
    frame=single,
    belowcaptionskip=2\baselineskip,
    tabsize=2,
}
\begin{lstlisting}[language=Java]
package d {
	public class d { ... }
}
package e {
	public class e { ... }
}
package f {
	public class f { ...}
}
\end{lstlisting}
\end{minipage}
\vspace{-0.01mm}

Listing~\ref{after_renaming} replaces the strings by a single character. Identifier mangling reduces the meta-information by replacing them with random alphabets. Random names do not carry any information about the object or the behavior. Hence, the interpretation becomes difficult. Proguard obfuscator within the Android SDK implements a similar approach. oBad Trojan leverages the identifier mangling with character permutations of ``o, O'', ``c, C'', ``i, I'' and ``l, L'' in lower and upper case~\cite{obad}. Listing~\ref{obad_snippet} illustrates the identifier renaming obfuscation. \newline

\definecolor{mygreen}{rgb}{0,0.6,0}
\lstset{
    captionpos=b,
    caption={oBad Trojan Identifier scrambling Obfuscation~\protect\cite{325schulz}.},
    label=obad_snippet,
    keywordstyle=\color{blue},
    commentstyle=\color{mygreen},
    basicstyle=\fontsize{7}{8}\selectfont\ttfamily,
    frame=single,
    numbers=left,
    belowcaptionskip=2\baselineskip,
    tabsize=2,
}
\begin{lstlisting}[language=Java]
public final class CcoCIcIf {
private static final byte [ ] COcocOlo;
private static boolean CcoCIcI;
private static long OoCOocll;
private static long OoCOocll;
private static String cOIcOOo;
private static final OoCOocll lOIlloc;
private static ArrayList occcclc;
private static final occccl coclClII;
private static Thread ooCclcC;
}
\end{lstlisting}

\subsection{PREVENTIVE TRANSFORMATIONS}
\indent Preventive transformation techniques are employed to evade the commercial debugging and de-compilation tools. Preventive obfuscation takes advantage weak or non-existent mapping between the high-level language and its corresponding bytecode to inject illegal, unused or rarely used bytecode. Changes in Java bytecode crashes the decompiler due to the presence of new instructions. Figure~\ref{fig:Obfuscation-Classification} illustrates the preventive obfuscation techniques.

%\vspace{-0.2cm}
\par \textbf{Anti Debugging:} This technique is a preventive obfuscation that inserts some validation code to identify the presence of a debugger. Once it identifies being executed in the debugger, the app behaves benign without revealing the malicious behavior.

\par \textbf{Anti De-compilation:} The transformation prevents reverse engineering of Dalvik or Java bytecode to high-level programming constructs. The obfuscator employs \texttt{goto} constructs valid with the bytecode which are not a part of Java language.
%\vspace{0.1cm}

%\vspace{-0.5cm}
\par \textbf{Bytecode Encryption:} Bytecode Encryption changes the structure of data. Low et al.~\cite{Low98javacontrol} suggest that encryption methods as an alternative to defeat decompilation. Code transformers encrypt the data with innovative methods to decrypt the encrypted content.

%\vspace{-0.2cm}
\par \textbf{Android Manifest Obfuscation:} Another technique to obfuscate data in Android application is Manifest Obfuscation. However, Android middleware verifies the manifest meta-data; some obfuscation techniques have been observed aiming at producing errors during the binary decoding~\cite{325schulz}. In the process of \texttt{APK} file creation, the manifest is compiled into a binary XML file. DexGuard~\cite{dexguard:2014}, the commercial extension of ProGuard~\cite{proguard} can obfuscate and optimize the binary XML files to avoid detection.
%\vspace{0.15 cm}

\par \textbf{String Encryption:} String Obfuscation hides the plain text strings that reveal sensitive information. The sensitive plaintext information can be misused. The strings must be available in plaintext form at runtime. Hence, the developer prefer to encrypt the plaintext string. String Obfuscation can be employed with an invertible encryption function employing \emph{AES}, \emph{DES} or \emph{XOR} encryption. Resultant output replaces the original plaintext. The process generates a byte sequence corresponding to each string as a data array stored in a private static class field. The strings are brought back to the original state at runtime. Hence, the strings are retrieved as plaintext when required. Dynamic analysis is useful as it extracts runtime information. Code listings~\ref{before_stringencryption} and ~\ref{after_stringencryption} illustrates code snippet before and after string encryption. The encrypted code converts readable string information into an array of strings with \texttt{ASCII} characters and obscures the code readability.

\begin{minipage}{.43\textwidth}
\definecolor{mygreen}{rgb}{0,0.6,0}
\lstset{
    captionpos=b,
    caption={before encryption~\protect\cite{defcon22}},
    label=before_stringencryption,
    keywordstyle=\color{blue},
    commentstyle=\color{mygreen},
    basicstyle=\fontsize{7}{8}\selectfont\ttfamily,
    frame=single,
    numbers=left,
    belowcaptionskip=2\baselineskip,
    tabsize=2,
}
\begin{lstlisting}[language=Java]
//original code
public void onClick(DialogInterface arg1, 
int arg2) { 
 try { Class.forName("java.lang.System")
.getMethod("exit", Integer.TYPE)
.invoke(null, Integer.valueOf(0)); 
 	return; 
 	} catch:(Throwable throwable) { 
 	throw throwable.getCause(); 
   } 
}
\end{lstlisting}

\end{minipage}\hfill
\begin{minipage}{.47\textwidth}
\definecolor{mygreen}{rgb}{0,0.6,0}
\lstset{
    captionpos=b,
    caption={string encrypted~\protect\cite{defcon22}},
    label=after_stringencryption,
    keywordstyle=\color{blue},
    commentstyle=\color{mygreen},
    basicstyle=\fontsize{7}{8}\selectfont\ttfamily,
    frame=single,
    belowcaptionskip=2\baselineskip,
    tabsize=2,
}
\begin{lstlisting}[language=Java]
//String encryption
public void onClick(DialogInterface arg1,
int arg2)
 { try { 
 	Class.forName(COn.ˊ(GCOn.ˋ[0xA], 
COn.ˋ[0x09], GCOn.ˋ[0xB]))
.getMethod(COn.ˊ(i1, i2, i2 | 6), 
Integer.TYPE) 
 	.invoke(null, Integer.valueOf(0)); 
	return; 
     } catch:(Throwable throwable) { 
 	throw throwable.getCause(); 
       } 
}

\end{lstlisting}
\end{minipage}
\vspace{0.2 cm}

\par
\begin{figure}[H]
    \centering
    \includegraphics[width=0.2\textwidth]{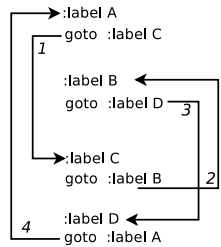}
    \caption{Bad code injection.}
    \label{fig:badcode-injection}
\end{figure}

 \textbf{Bad Code Injection:} Figure~\ref{fig:badcode-injection} illustrates recursive \texttt{goto} sequences with an indirect recursion. This transformation thwarts the existing static analysis tools (disassemblers, decompilers) by inserting bytecode statements not available with high-level language. 

Some instructions may be valid in Java. However, their corresponding Dalvik instructions may not be available. For example, Java language does not have \texttt{goto} construct. However, when a loop or switch constructs are converted to Dalvik, the \texttt{goto} statement is generated. Hence, \texttt{goto} can be injected in the Dalvik bytecode. In particular, a dummy method with recursive \texttt{goto} statements can be used. 

\vspace{0.2cm}
\par \textbf{Reflection:} Reflection is a powerful Java programming technique to extend additional functionality such as verifying backward compatibility or dynamically load methods. It is used in debugging and testing tools~\cite{Wognsen201425}. Java Reflection application programming interface (API) allows a program to access the class information during execution to~\cite{Wognsen,Wognsen201425}: (1) create new objects; (2) invoke a method; or (3) modify the code control-flow. Reflection can be alternatively employed as data obfuscation technique. Code listings~\ref{without_reflection} and~\ref{reflection_obfuscation} illustrate the importance of reflection obfuscation. \newline

\vspace{0.2cm}
\begin{minipage}{.45\textwidth}
\definecolor{mygreen}{rgb}{0,0.6,0}
\lstset{
    captionpos=b,
    caption={before obfuscation},
    label=without_reflection,
    keywordstyle=\color{blue},
    commentstyle=\color{mygreen},
    basicstyle=\fontsize{7}{8}\selectfont\ttfamily,
    frame=single,
    numbers=left,
    belowcaptionskip=2\baselineskip,
    tabsize=2,
}
\begin{lstlisting}[language=Java]
// A Standard Call in Java
Crypto cryptModule = new Crypto();
privateKey = cryptModule.getPrivateKey();   
\end{lstlisting}

\end{minipage}\hfill
\begin{minipage}{.47\textwidth}
\definecolor{mygreen}{rgb}{0,0.6,0}
\lstset{
    captionpos=b,
    caption={code reflection obfuscation},
    label=reflection_obfuscation,
    keywordstyle=\color{blue},
    commentstyle=\color{mygreen},
    basicstyle=\fontsize{7}{8}\selectfont\ttfamily,
    frame=single,
    belowcaptionskip=2\baselineskip,
    tabsize=2,
}
\begin{lstlisting}[language=Java]
// Invoke using reflection  
Object reflectedClassInstance = 
Class.forName (pe.mnit.secureApp.Crypto).
newInstance();

Method methodToReflect =
reflectedClassInstance.getClass().
getMethod(getPrivateKey);

Object invokeResult = methodToReflect.invoke
(reflectedClassInstance);

\end{lstlisting}
\end{minipage}
\vspace{-0.01mm}

Reflections is used in Android for a variety of purposes such as:
    \begin{enumerate}
        \item \textbf{Invoking hidden API methods: }Certain Android framework features are intentionally hidden from the developers during compilation. The hidden API may not support all Android devices, or the stable version is yet to be made public. 
\vspace{0.1cm}
        \item \textbf{Providing backward compatibility: }New Android versions incorporate additional features and get released in the incremental higher versions. A developer may use reflection to verify if a particular feature present in older version exists with the new version. If available, then only calls that feature/method.
\vspace{0.1cm}
        \item \textbf{Interacting with JSON data: }Data from server is loaded in JavaScript Object Notation format (\emph{JSON}) from the web. Hence, it is parsed using reflections in Android. 
\vspace{0.1cm}
        \item \textbf{Libraries: }Native libraries can be loaded in an APK using reflection API. Some applications employ custom native libraries for improved performance.

    \end{enumerate}

\subsection{Repackaging Popular Apps}
\par Repackaging Android \texttt{APK} is a popular practice among the malware authors. In particular, malware authors reverse engineer the popular apps. The app is reverse engineering, malware payload is inserted and released at less monitored Android third-party app markets. In~\cite{semantic:2013}, authors discuss a set of methods to replace the app developer library by the plagiarist ad-libraries to divert the advertisement revenues. Thus, a developer is robbed off the advertisement revenue.  

\indent Repackaging is the one of the widely used Obfuscation technique employed by malware authors on Android platform~\cite{conf/mobisys/GiblerSCCZC13}. Zhou et al.~\cite{zhou:dissecting,Zhou:2013:AWA:2484313.2484315} reported 86\% repackaged malware among the 1260 Android Malware Genome dataset. Another study reported 5-13\% repackaged and malicious applications among the well-known six third-party app stores~\cite{zhou:dissecting}. The other third-party app stores do not have robust app verification and vetting. Hence, there is a higher chance of plagiarized and repackaged apps. Furthermore, authors in~\cite{conf/mobisys/GiblerSCCZC13} reported around 30\% plagiarized, and cloned apps even at the Google Play. The repackaging process typically employs following steps:

%\begin{figure*}[h!]
%    \centering
%    \includegraphics[scale=0.33]{figures/repackaging}
%    \caption{\label{repackaging} APK Repackaging procedure.}
%\end{figure*}

\begin{itemize} 
\item Download the popular app from the Play store.
\item Disassemble the app with \textit{apktool}~\cite{apktool:2012}.
\item Develop malicious payload either in Dalvik bytecode or Java source~\cite{dxtool}.
\item Add the malicious payload inside a popular app.
\item Modify the \texttt{AndroidManifest.xml} and/or \textit{resources} if required.
\item Assemble modified source with \textit{apktool}.
\item Distribute repackaged app by with another certificate to the less monitored third-party markets.
\end{itemize}
%\vspace{0.2cm}
\par Repackaging is used by malware authors as a technique to evade commercial anti-malware. Repackaged apps create an imbalance in-app distribution markets, hurt the developer reputation, and inflict monetary loss to the developers~\cite{6999911,DBLP:conf/trust/HuangZLW13}. Malware authors also employ repacking to divert the advertisement revenues by replacing the original advertisements with their own. AndroRAT APK Binder~\cite{androrat} repackages and generates trojanized version of popular, legitimate app appending remote access functionality. The adversary can remotely force the infected device to send premium-rate SMS, make voice calls, access the device location, record video and audio and access the device files without the device user knowledge.

\subsection{CUSTOM OBFUSCATION TECHNIQUES}

\indent Malware authors have been very active in developing customized obfuscation to defeat the anti-malware. In this section, we discuss following custom techniques primarily used by malware writers to protect the malicious code. Apvrille et al.~\cite{vb:2014} list some interesting techniques used by malware authors to defeat application analysis and reverse engineering:

\par \textbf{Using very long class names:} Decompilers tend to crash when the class names are too long or written in \texttt{non-ASCII} format~\cite{schulz:codeprotection}. Few malicious apps have demonstrated use of very long class names to defeat reverse engineering tools \texttt{Android/Mseg.A!tr.spy}~\cite{Strazzere:2012} reported in the wild has successfully evaded commercial anti-malware with this technique. 

\par \textbf{Hide Packages, JAR inside raw resources:} Malware developers hide the malicious executable package inside the resource files to avoid code inspection. For example, \texttt{Android/SmsZombie.A!tr} hides malicious package within a jpeg file \texttt{a33.jpg}, in the assets directory~\cite{blackhat:2014}. \texttt{Android/Gamex.A!tr} conceal an encrypted malicious package within asset \texttt{logos.png}, again an image file~\cite{blackhat:2014}. Table \ref{table:example-hiding-malicious} enlists interesting malicious apps using the discussed techniques~\cite{vb:2014}.

\par \textbf{NOP to modify bytecode control flow:} No Operation (\texttt{NOP}) is an assembly language instruction, that does nothing at all. A sequence of \texttt{NOP} instructions wastes the CPU cycles and adds to the code complexity. Malware authors can leverage \texttt{NOP} instruction to modify the bytecode flow to hide the actual control flow of the program. Inserting the \texttt{NOP} instruction changes the syntax structure, a common technique to evade anti-malware. This approach is easy and used by quite a few malware authors.

\par \textbf{Path Obfuscation:} Path obfuscation is used to achieve the cloning transformation~\cite{securityfocus}. The idea is to change the path such that there are different methods, but the same meaning. This technique is used within URLs to obfuscate the HTTP-based attacks~\cite{securityfocus}.

\par
\begin{figure}[H]
    \centering
    \includegraphics[width=0.65\textwidth]{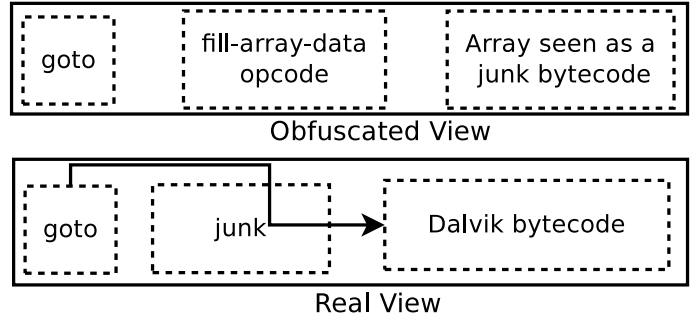}
    \caption{Hiding bytecode in the array of fill-array-data.}
    \label{fig:Hiding-bytecode}
\end{figure}

 \textbf{Hiding bytecode:} Hiding bytecode obfuscates \texttt{APK} with a variable length fill-array-data-payload instruction to hide the original bytecode~\cite{Schulz:2012}. This technique can be detected by looking for Dalvik bytecode employing \texttt{goto} obfuscation followed by fill-array-data opcode, illustrated in Figure~\ref{fig:Hiding-bytecode}. The bytecode is hidden in the fill-array-data which remains invisible to the disassemblers.

%\vspace{-0.3cm}

\begin{table}[h!]
\footnotesize
%\rowcolors{1}{white}{lightgray}
\centering
\scalebox{1}{
    \begin{tabular}{| l  | p{8.1cm} | }
    \hline
\textbf{Android malware name} & \textbf{Purpose of Obfuscation}  \\ \hline
\texttt{Gamex.A!tr}&   The asset \texttt{log-os.png} is a ZIP. However, it has the capability of being a valid ZIP file (for instance, when XOR'ed with the right key). \\ \hline
\texttt{SmsZombie.A!tr} &   Hides malicious package in \texttt{a33.jpg}.\\ \hline
\texttt{DroidCoupon.A!tr} &    The Rage Against the Cage (root exploit, commonly used to root Android phones) is hidden inside a png image in the resources.\\ \hline

    \end{tabular}
}    
    \caption{hidden malware payload inside \texttt{APK} resource.}
    \label{table:example-hiding-malicious}
\end{table}
%\vspace{-0.6cm}

\vspace{-0.4cm}
\par \textbf{String table:} A string table can be used to hide strings. In this technique, a malicious app builds a string table as an array of characters. The table hides suspicious strings. The reverse engineering tools fails to identify strings. Listing~\ref{stringtable_snippet} illustrate use of string table as an obfuscation technique.
%\end{enumerate}

\definecolor{mygreen}{rgb}{0,0.6,0}
\lstset{
    captionpos=b,
    caption={String Table},
    label=stringtable_snippet,
    keywordstyle=\color{blue},
    commentstyle=\color{mygreen},
     basicstyle=\fontsize{7}{8}\selectfont\ttfamily, 
    frame=single,
    numbers=left,
    belowcaptionskip=2\baselineskip,
    tabsize=2,
}
\begin{lstlisting}[language=Java]
package Eg9Vk5Jan;
class x18nAzukp {
   final private static char[][] OGqHAYqtswt8g;
   static x18nAzukp()
   { v0 = new char[][48];
     v1 = new char[49];
     v1 = {97, 0, 110, 0, 100, 0, 114, 0, 111, ...
     v0[0] = v1;}
protected static String rLGAEh9gGn73A(int p2) {
     return new String(Eg9Vk5Jan.x18nAzukp.
OGqHAYqtswt8g[p2]);
} ...
new StringBuilder(x18nAzukp.rLGAEgGn73A(43))
\end{lstlisting}
\vspace{-0.01cm}

\begin{figure*}
    \centering
    \includegraphics[scale=0.3]{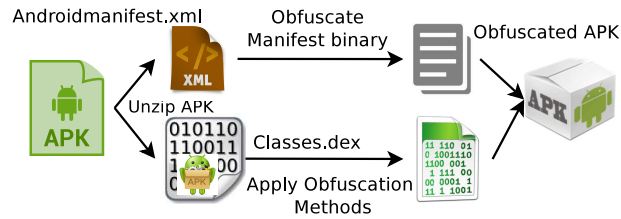}
    \caption{\label{obfuscationlayout} \texttt{APK} Obfuscation and optimization methodology.}
\end{figure*}

\section{Custom Code Obfuscation Tools}
\label{sec_formalwareauthors}
%\vspace{-0.1cm}
\noindent Obfuscation techniques, though not invincible, are very popular among malware writers. Figure~\ref{obfuscationlayout} illustrates the steps employed by obfuscation methods to transform and optimize the code.

\vspace{-.45cm}

\subsection{\emph{Proguard \protect\footnote{http://proguard.sourceforge.net (accessed August, 2016.)}}} 
%\vspace{-0.3cm}
Proguard obfuscator is a part of Android software development kit (SDK)~\cite{proguard}. Proguard is a Java source code transformer. Google recommends Proguard to protect Android \texttt{APK}. Proguard has an in-built optimizer, shrinker and a weak obfuscator. The Obfuscator ``tool removes unused or unnecessary code, merges the identical code blocks, employs \texttt{peep hole} optimization, removes debug information, renames objects and restructures the original code''~\cite{defcon22}. %Figure~\ref{obfuscationlayout} illustrates the \texttt{APK} obfuscation and optimization. Proguard renames the relevant paths, class names, methods and Java identifiers with irrelevant alphabets. The original variables cannot be recovered once replaced with the Proguarded identifiers. Examples of this behavior could be pointed out in Listings~\ref{before_proguard} and ~\ref{after_proguard} for completeness purposes. However, the transformed samples are easy to reverse engineer due to the availability of the original code layout. Proguard has following notable features: (i) Bytecode optimization; (ii) identifier renaming; (iii) reduces \texttt{classes.dex} size. Code listings~\ref{before_proguard} and~\ref{after_proguard} illustrate the normal and Proguard obfuscated code, respectively. 

\begin{minipage}{.47\textwidth}
\definecolor{mygreen}{rgb}{0,0.6,0}
\lstset{
    captionpos=b,
    caption={before obfuscation~\protect\cite{defcon22}},
    label=before_proguard,
    keywordstyle=\color{blue},
    commentstyle=\color{mygreen},
    basicstyle=\fontsize{7}{8}\selectfont\ttfamily,
    frame=single,
    numbers=left,
    belowcaptionskip=2\baselineskip,
    tabsize=2,
}
\begin{lstlisting}[language=Java]
//original code listing
public static String exec(String cmd, 
Boolean root) { 
        BufferReader mybufferReader; 
        DataStream   testdataOutputStream; 
        Process process; 
        String string = "sh"; 
        if(root.booleanValue()) { 
            string = "su"; 
  	} 
     StringBuild teststringBuilder = 
new StringBuilder(); 
        try { 
        process = Runtime.getRuntime()
.exec(string); 
 dataOutputStream = new DataOutputStream
(process.getOutputStream());
dataOutputStream.writeBytes(cmd + "\n"); 
mybufferReader = new BufferedReader( 
new InputStreamReader(process
.getInputStream())); 
        } 

\end{lstlisting}

\end{minipage}\hfill
\begin{minipage}{.44\textwidth}
\definecolor{mygreen}{rgb}{0,0.6,0}
\lstset{
    captionpos=b,
    caption={Proguarded code~\protect\cite{defcon22}},
    label=after_proguard,
    keywordstyle=\color{blue},
    commentstyle=\color{mygreen},
    basicstyle=\fontsize{7}{8}\selectfont\ttfamily,
    frame=single,
    belowcaptionskip=2\baselineskip,
    tabsize=2,
}
\begin{lstlisting}[language=Java]
// Proguarded output

public static String a(String arg6, 
Boolean arg7) { 
        Process process; 
        String string = "mksh"; 
        if(arg7.booleanValue()) { 
            string = "su"; 
        } 
      
  testStringBuilder stringBuilder = 
new StringBuilder(); 
        try { 

            process = Runtime.getRuntime()
.exec(string); 

       DataOutputStream dataOutputStream = 
new DataOutputStream(process
.getOutputStream()); 
            dataOutputStream

.writeBytes(String.valueOf(arg6) + "\n"); 

       BufferedReader bufferedReader = 
new BufferedReader( 
 	}

\end{lstlisting}
\end{minipage}

\subsection{\textbf{Allatori}\protect\footnote{http://www.allatori.com  (accessed August, 2016.)}} Allatori~\cite{allatori:2013} is a commercial product from Smardec. Besides identifier renaming, Allatori tool can also modify the source code. Allatori is a code optimizer, shrinker, obfuscator and a watermarking tool. The tool obscures the loops within the program such that reverse engineering tools are easily evaded. Such an approach increases the code size and makes the program logic less readable. Moreover, Allatori also encrypts the strings and decrypts them at runtime. Allatori has the following notable features: (i) reduced \texttt{.dex} file size; (ii) improves \texttt{APK} execution speed; (iii) decreases memory usage; (iv) removes debug code; and (v) employs simple obfuscation.

%\vspace{-0.5cm}
\subsection{\textbf{DexGuard}\protect\footnote{ https://www.guardsquare.com/dexguard (accessed August, 2016.)}} DexGuard~\cite{dexguard:2014} is a professional code optimizer and obfuscator developed by Eric Lafortune. It performs code optimization, code shrinking, and encryption. Dexguard converts the class and methods names into \texttt{non-ASCII} values and strings are encrypted with the encryption algorithms. DexGuard has following features in addition to Proguard: (i) Reflection obfuscation at runtime; (ii) Encrypt strings within an array; (iii) assets, resource and library encryption; (iv) encrypts Java class names; and (v) identifies \texttt{APK} tampering.

\noindent\begin{minipage}{.40\textwidth}
\definecolor{mygreen}{rgb}{0,0.6,0}
\lstset{
    captionpos=b,
    caption={before obfuscation},
    label=before_dexguard,
    keywordstyle=\color{blue},
    commentstyle=\color{mygreen},
    basicstyle=\fontsize{7}{8}\selectfont\ttfamily,
    frame=single,
    numbers=left,
    belowcaptionskip=2\baselineskip,
    tabsize=2,
}
\begin{lstlisting}[language=Java]
// original code

public void onClick(
DialogInterface arg3, int arg4) 
{ 
	System.exit(0); 

}																										.	
\end{lstlisting}

\end{minipage}\hfill
\begin{minipage}{.52\textwidth}
\definecolor{mygreen}{rgb}{0,0.6,0}
\lstset{
    captionpos=b,
    caption={Dexguarded~\protect\cite{defcon22}},
    label=after_dexguard,
    keywordstyle=\color{blue},
    commentstyle=\color{mygreen},
    basicstyle=\fontsize{7}{8}\selectfont\ttfamily,
    frame=single,
    belowcaptionskip=2\baselineskip,
    tabsize=2,
}
\begin{lstlisting}[language=Java]
// Dexguard obfuscated code
public void onClick(
DialogInterface arg7, int arg8) { 
   try { 
 	
	Class.forName
("java.lang.System").getMethod("exit",Integer.
TYPE).invoke(null, Integer.valueOf(0)); 
	return; 
   } catch:(Throwable throwable) { 
 	throw throwable.getCause(); 
     } 
}
\end{lstlisting}
\end{minipage}

\subsection{\textbf{dalvik-obfuscator}\protect\footnote{https://github.com/thuxnder/dalvik-obfuscator  (accessed August, 2016.)}} dalvik-obfuscator is an open-source bytecode transformation tool~\cite{Dalvikobfuscator}. The analyst must provide an \texttt{APK} file as input to obtain the obfuscated app version. Dalvik-obfuscator employs the popular junk byte injection approach on the x86 platform. Dalvik-obfuscator is composed of a set of tools/scripts to obfuscate and manipulate \texttt{.dex} files. The obfuscator iterates through all the methods, insert junk bytes and unconditional branch in the code block, to ensure it is never executed.

\subsection{\textbf{APKfuscator}\protect\footnote{https://github.com/strazzere/APKfuscator (accessed August, 2016.)}} APKfuscator is a dead code injection obfuscator~\cite{apkfuscator}. Available as an open-source, APKfuscator employs quite a few variations of dead code injection. APKfuscator functions on bytecode level and leverages the Unix filesystem restriction that does not allow a class name to exceed 255 characters. APKfuscator employs three variations of dead code injection: (i) insert illegal opcodes; (ii) use legitimate opcodes into ``bad'' objects; and (iii) inject code inside the \texttt{.dex} header by exploiting a discrepancy between the claims of the official \texttt{.dex} documentation and \texttt{DEX} verifier.

\section{CODE PACKERS AND PROTECTORS} 
\label{sec_protectorspackers}

Android code Packers insert new malicious \texttt{DEX} file and encrypt the \texttt{classes.dex} in the existing \texttt{.dex} file within an \texttt{APK}. The new \texttt{.dex} is decrypted in memory during runtime using \texttt{DexClassLoader}, a Java class loader~\cite{artvsdalvik,vb:2014,bangcle}. Packers were developed for Android platform to protect the legitimate app from unwanted tampering and modifications. However, malware developers use packers to obfuscate the dalvik bytecode and evade anti-malware scanners.

\subsection{Code Packers}
\begin{figure}[H]
    \centering
    \includegraphics[scale=0.25]{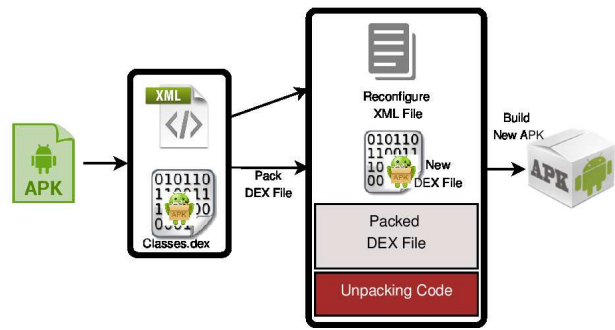}
    \caption{\label{packerdemo} Code Packing steps.}
\end{figure}

Packing encrypts the executable code to prevent static analysis. The unpacker routine brings the code in readable form. Malware developers employ executable code packers to evade reverse engineering of malicious \texttt{DEX}. The runtime unpacking routine brings the code into its original form. The code protectors can also be used in conjunction with existing obfuscation techniques to harden static analysis. Figure~\ref{packerdemo} illustrates working of code packers.

%\begin{enumerate}
\vspace{-0.3cm}
\par \textbf{APK Protect\protect\footnote{https://sourceforge.net/projects/apkprotect/ (accessed August, 2016.)}:} is a professional code packing tool with anti-debug, anti-decompile and anti-disassembly support~\cite{apkprotect}. The packer performs code obfuscation with string encryption with Base64 encoding. It employs Java reflections to load the code dynamically. APK Protect has following features: (i) debugger detection; (ii) app encryption; (iii) code reflections; (iv) anti-debugging; (v) anti-disassembly; and (vi) anti-decompilation. 

\vspace{0.2cm}
\par
\begin{figure}[H]
    \centering
    \includegraphics[scale=0.35]{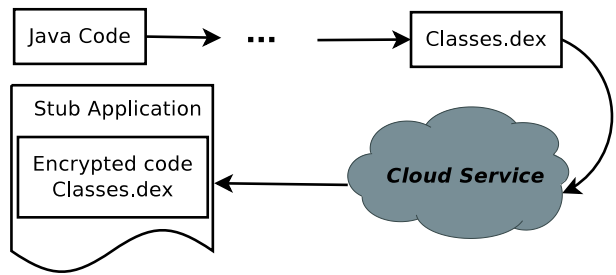}
    \caption{\label{hosedex2jar} HoseDex2Jar Packer.}
\end{figure}

 \textbf{HoseDex2Jar\protect\footnote{
                  https://github.com/strazzere/dehoser (accessed August, 2016.) }:} is an executable packer to encrypt \texttt{.dex}, repackage encrypted file, and store inside 112 header bytes of the target \texttt{.dex}. Figure~\ref{hosedex2jar} illustrate code packing procedure with HoseDex2Jar. The code packer performs: (i) \texttt{.dex} repackaging; and (ii) code encryption.

\vspace{-0.4cm}

\par
\begin{figure}[H]
    \centering
    \includegraphics[scale=0.35]{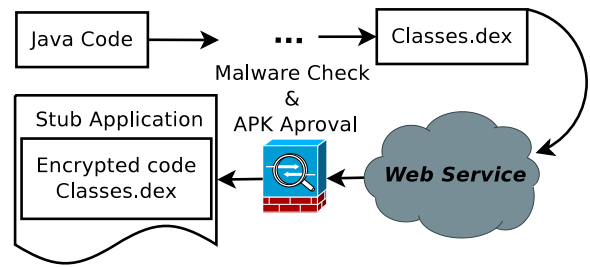}
    \caption{\label{bangclepacker} Functioning of Bangcle Packer. }
\end{figure}

 \textbf{Bangcle:}\protect\footnote{https://www.crunchbase.com/organization/bangbang-security/entity (accessed August, 2016.):} is an online \texttt{APK} packing tool~\cite{bangcle}. The developer must register at the Bangcle and use Bangcle Assistant tool to upload the package. The app must be uploaded with the Keystore to retrieve the protected \texttt{APK}. The packing process changes the \texttt{APK} name, inserts new assets, native libraries and modifies the Android manifest. Figure~\ref{bangclepacker} illustrates the code packing procedure. Bangcle packer provides (i) online \texttt{APK} wrapping; (ii) resists reverse engineering; (iii) online anti-debugging, anti-tamper and anti-decompilation service.

%\vspace{-0.8cm}

\vspace{0.2cm}
\par
\begin{figure}[H]
    \centering
    \includegraphics[scale=0.3]{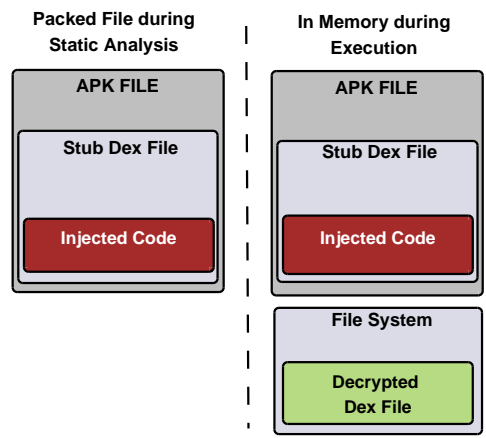}
    \caption{\label{unpacking} Unpacking procedure during Dynamic analysis.}
\end{figure}

 \textbf{PANGXIE\protect\footnote{
                  http://www.packers.com/ (accessed August, 2016.) }:} is a Proof-of-Concept (PoC) packer armed with anti-debugging and anti-tampering techniques~\cite{defcon22}. The packer encrypts the \texttt{.dex} bundled inside the assets of an \texttt{APK}. Though, PANGXIE evades static analysis, the obfuscator increases the APK size. Figure~\ref{unpacking} illustrates the code unpacking procedure based on runtime execution.

%\vspace{-0.4cm}

\subsection{Comparison of Obfuscation and Protection Techniques}
In the following, we evaluate the effectiveness of Packers based on following attributes illustrated in Table~\ref{packers-comparison}.

\par \textbf{Code Obfuscation:} This technique prevents the analysis of the code either at source code or bytecode level~\cite{Collberg02watermarking}. The Android Integrated Development Environment (IDE) has Proguard, an in-built obfuscator to transform the Java class names, fields, and method names \cite{proguard}. Additionally, the persistent methods like control-flow obfuscation, reorder program flow, readable string encryption, and dynamic code loading has been employed by recent malware~\cite{rcj14,adam:2013}. Furthermore, the app developers also use Java Reflection methods and invoke the native code functionality using Java Native Interface (JNI) to hinder static analysis. 

\par \textbf{Dynamic Code Modification:} Android user apps, developed in Java are converted to Dalvik bytecode using~\texttt{Dx} tool~\cite{artvsdalvik,LinuxAndroidTools}, part of Android SDK. Dalvik Virtual Machine (DVM) verifies the bytecode, and executes it in the VM. It is difficult for a programmer to modify bytecode from VM during runtime. Malware developers use Java Native Interface (JNI) methods to modify and load the bytecode at runtime DVM~\cite{DBLP:conf/dsn/QianLSC14}. ART pre-compiles the \texttt{.dex} file as \texttt{oat} in the ELF format. To avoid precompilation, the native code modifies the instructions of \texttt{.dex} and \texttt{.oat} data structures. 

\par \textbf{Dynamic Code Loading:} Android permits loading of external \texttt{.jar} or a \texttt{.dex} at runtime. The executable code appears different compared to its static visuals. Malware authors leverage this facility by encrypting the malicious executable, then decrypting and loading them in the VM at runtime. 

\par \textbf{Anti-debugging:} Android kernel has in-built GNU debugger \texttt{(gdb)} to attach process for debugging. Packers get attached to \texttt{Ptrace}~\cite{LinuxAndroidTools} tool to evade gdb under the assumption that only a single process can be attached to the target process for monitoring. Hence, the gdb cannot attach itself to the \texttt{APK}; preventing the \texttt{APK} debugging. Advanced Packers identify Java Debug Wire Protocol (JDWP) threads being attached to an \texttt{APK}. Furthermore, the Packers can identify themselves being monitored within emulated environment. 

Table~\ref{packers-comparison} illustrates a comparison of the crucial methods prevalent among the leading Packers based upon their properties. The Android Packers are a new phenomenon and evolving quickly. We examined known malicious apps on portals and packed with stand alone Packers during October and November 2015. Table~\ref{packers-comparison} shows that all of the Packers use one or more code obfuscation techniques and append shared library. The comparison shows that few Packers support ART. Except Bangcle and Ijiami, other packers are not equipped to perform runtime modifications.

\begin{table*}[h!]
\footnotesize
%\rowcolors{1}{white}{lightgray}
\centering
%\resizebox{\textwidth}{!}{%
\scalebox{0.85}{
\begin{tabular}{@{}lcccccccc@{}}
\toprule
\textbf{} & \multicolumn{8}{c}{\textbf{Packer Protection Techniques}} \\ \midrule
\makecell{Packer} & \makecell{Code\\ Obfuscation} & \makecell{Dynamic\\ Code \\ Loading} & \makecell{Dynamic\\ Code \\ Modification} & \makecell{Debugger\\ Detection} & \makecell{Append\\ shared \\ Libraries} & \makecell{Additional\\ Class\\ insertion} & \makecell{DVM\\ Support} & \makecell{ART\\ Support} \\ \midrule
APKProtect & \ding{52} & \ding{52} & \ding{52} & \ding{52} & \ding{52} & \ding{52} & \ding{52} & \ding{55} \\ \midrule
Ali & \ding{52} & \ding{52} & \ding{52} & \ding{52} & \ding{52} & \ding{52} & \ding{52} & \ding{55} \\ \midrule
Baidu & \ding{52} & \ding{52} & \ding{52} & \ding{52} & \ding{52} & \ding{52} & \ding{52} & \ding{52} \\ \midrule
Bangle & \ding{52} & \ding{52} & \ding{55} & \ding{52} & \ding{52} & \ding{52} & \ding{52} & \ding{52} \\ \midrule
Ijiami & \ding{52} & \ding{52} & \ding{55} & \ding{52} & \ding{52} & \ding{52} & \ding{52} & \ding{52} \\ \midrule
HoseDex2jar & \ding{52} & \ding{52} & \ding{55} & \ding{52} & \ding{55} & \ding{55} & \ding{52} & \ding{55} \\ \midrule
Pangxie & \ding{52} & \ding{55} & \ding{55} & \ding{55} & \ding{55} & \ding{52} & \ding{52} & \ding{55} \\ \bottomrule
\end{tabular}
}
\caption{Comparing packer protection methods}
\label{packers-comparison}
\end{table*}

\section{ANDROID BYTECODE ANALYSIS}
\label{sec_bytecodetools}

\begin{figure*}[h!]
    \centering
    \includegraphics[width=1\textwidth]{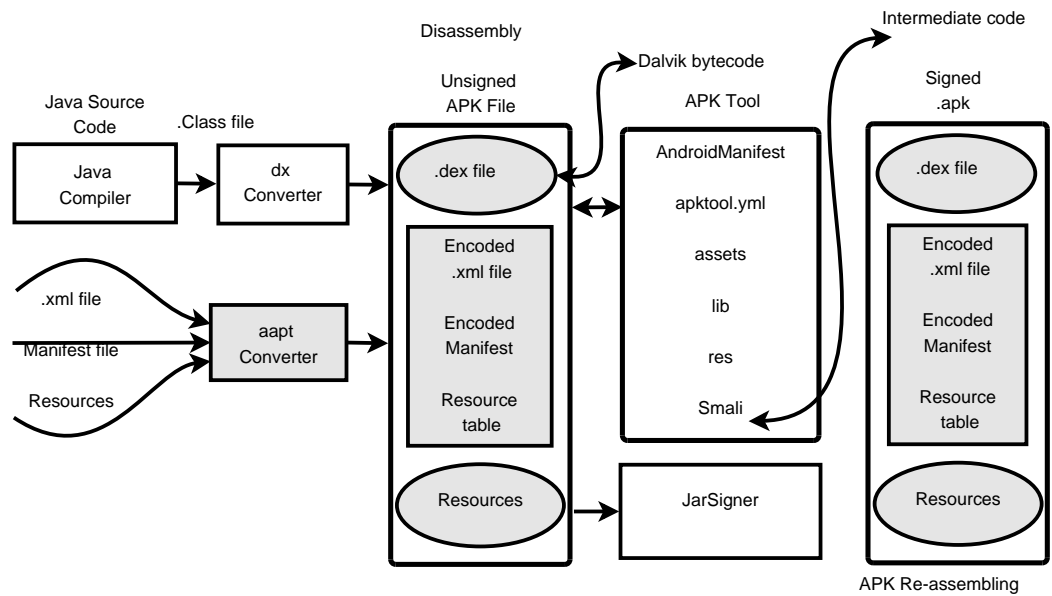}
    \caption{Disassembly of an \texttt{APK} file}
    \label{fig:disassembly}
\end{figure*}

\indent In this section, we discuss various Dalvik bytecode de-obfuscation tools and techniques. To de-obfuscate an \texttt{APK}, the reverse engineered bytecode must be available. Hence, we discuss the Dalvik bytecode, as it is the nearest readable intermediate code for disassembly and analysis~\cite{Faruki:trustcom}. Figure~\ref{fig:disassembly} illustrates the process of bytecode extraction and analysis of a normal Android \texttt{APK} file. This process is followed by analysts and malware authors for their respective purpose.

\vspace{-0.1cm}
\subsection{Bytecode and De-Obfuscation Tools}
In the following, we briefly discuss open-source and commercial de-obfuscation tools.

\par \textbf{Dexdump:} Dexdump is a part of the Android software development kit (SDK)~\cite{dexdump:2013}. Dexdump is a Dalvik executable (\texttt{dex}) file dissection tool that can be used to disassemble Dalvik bytecode. Dexdump is a linear sweep disassembler that finds a valid instruction at the last byte of the instruction being analyzed. Linear sweep disassembler can be circumvented easily by inserting the junk bytes to prevent the disassembly.

%\vspace{0.15cm}
\par \textbf{Smali:} Smali is a Dalvik bytecode assembler~\cite{smali:2013}. The package contains baksmali to disassemble the assembled code. Hence, both smali and baksmali can be used to disassemble, modify and reassemble the Android apps~\cite{JesusFrekesmali}. Baksmali performs recursive disassembly by following the address of jump towards the current instruction.

%\vspace{0.15cm}
\par \textbf{Androguard:} Desnos proposed Androguard, an open-source static analysis tool to reverse engineer the \texttt{APK} files~\cite{androguard:2013}. Androguard has a recursive disassembler and semantic analysis methods to identify similarity, dissimilarity, call graph analysis and signature of malicious apps. It provides a graphical preview of call graphs to assist the human analyst to detect cloned and repackaged APK files.

%\vspace{0.15cm}
\par \textbf{APKInspector:} APKInspector is static, bytecode analysis tool. APKInspector leverages Ded~\cite{ded}, Smali/Baksmali~\cite{smali:2013}, APKtool~\cite{apktool:2012} and Androguard~\cite{androguard:2013} to reverse engineer APK bytecode. APKInspector performs: (i) meta-data analysis; (ii) sensitive permission usage; (iii) generates bytecode control-flow graph; (iii) generates call-graph with call-in and call-out structures.

%\vspace{0.15cm}
\par \textbf{dex2jar:} dex2jar is a disassembler to parse both the \texttt{dex} and \texttt{optimized dex} files, providing a lightweight API to access~\cite{Dextojar:2012}. dex2jar can also convert \texttt{dex} to a \texttt{jar} file, by re-targeting the Dalvik bytecode into Java bytecode. Moreover, it is possible to re-assemble the \texttt{jar} into a \texttt{dex} after modifications.

\par \textbf{Dare:} Dare project aims at re-targeting Dalvik bytecode within \texttt{classes.dex} to traditional \texttt{.class} files using strong type inference algorithm~\cite{dare}. This \texttt{.class} files can be further analyzed using a range of traditional techniques developed for Java applications, including the de-compilers. Octeau et al.~\cite{octeau2012retargeting} demonstrated that Dare is 40\% more accurate than dex2jar.

\par \textbf{Dedexer:} Dedexer tool~\cite{dedexer} disassembles the \texttt{classes.dex} into Jasmin, an intermediate code format. Furthermore, it creates a separate file for each class to maintain the package directory structure for easy readability~\cite{Wognsen,Wognsen201425}. However, unlike the apktool, it cannot re-assemble the intermediate disassembled class files.

%\vspace{0.2 cm}
\par \textbf{JEB:} JEB~\cite{jeb} is a leading professional Android reverse-engineering software available on Windows, Linux, and Macintosh. The GUI interactive decompiler analyzes reversed malware app content. App information such as manifest, resources, certificates, literal strings can be examined in Java source by providing an easy navigation through the cross-references. JEB converts Dalvik bytecode to Java utilizing Dalvik bytecode semantics. JEB has de-obfuscator for obfuscated bytecode ~\cite{Dextojar:2012,apktool:2012}. JEB supports plugins by allowing access to the decompiled Java code Abstract Syntax Tree (AST) through API. The feature is useful to perform custom analysis.
%\vspace{0.2 cm}

\par \textbf{IDA Pro:} IDA Pro~\cite{idapro:2013} is a commercial recursive disassembler popular for \texttt{x86} platform. The IDA 6.0 onwards supports the Android Dalvik bytecode disassembly. IDA has a complete GUI with options to extend the functionality with supported API plugins to extend analysis functionality. IDA Pro has an additional capability to disassemble specific parts of the code in a file selected by the user. Dalvik bytecode can be represented as a graph making it easy to follow the control flow within a program. 
%\vspace{0.1 cm}

\par \textbf{Dexter:} Dexter is a free online analysis service~\cite{dexter:2014} and static analysis tool to process an input \texttt{APK} file. It provides rich information about: (i) \texttt{APK} permission; (ii) obfuscated code and packages; and (iv) mapping between broadcast receivers and the data-store.

%\vspace{0.2 cm}
\par \textbf{Dexguard:} Dexguard is a set of scripts to perform automated strings de-obfuscation and recovery of the \texttt{dex} file~\cite{dexguard:2013}. This tool is a mix of static and dynamic analysis consisting of (i) \texttt{dex} File reader; (ii) Dalvik disassembler; (iii) Dalvik emulator; and (iv) \texttt{dex} File parser.

\vspace{0.2cm}

\par \textbf{Radare2:} Radare2~\cite{radare2} is an interactive bytecode disassembly and analysis tool with precise control during reverse engineering procedure. Radare2 decompiles the code using open-source decompiler boomerang~\cite{boomerang}. This tool is a recursive disassembler that allows a user to specify starting address for decompilation. The hybrid approach makes Radare2 more efficient against smart obfuscation techniques in comparison to other approaches.

%\end{itemize}

\subsection{Stealth Obfuscation} 
\label{section_futureofobfs}

Malware writers employ different code protection techniques to delay the malware being reverse engineered. Android malware is following the footsteps of Windows malware evolution with the recent addition of code packing tools. The Packers are used to generate unseen variants of known malware apps, further increasing the pressure on signature-based commercial anti-malware. In this section, we discuss the techniques that are possibly useful to hide the malicious code. Bremere et al. in~\cite{bremer:2013} demonstrated the possibility of bytecode injection inside any class having a virtual function. Such techniques can be used by attackers to serve the evil intents like:
%\vspace{-0.3cm}

\begin{enumerate}
\item To develop a benign app capable of bytecode interpretation and loading.
\item To read the bytecode from \texttt{APK} or a remote host.
\item To inject the malicious bytecode inside a benign app.
\end{enumerate}

%\vspace{-0.3cm}
At present, this technique is limited to return integer values. However, malware authors can misuse the extended version to inject malicious bytecode at runtime. Authors in~\cite{apvrille:2013} demonstrated a Proof-of-Concept code to hide the Dalvik methods against the reverse engineering tools. Furthermore, they presented the methods to detect such hidden methods. Malware authors can place the malicious logic within such hidden methods and evade anti-malware. However, the authors also developed the \texttt{Hidex} tool~\cite{apvrillhideseek} to detect invisible methods.

%% Last lines to be changed in this one %%%%%
In~\cite{blackhat:2014}, the authors developed several methods capable of manipulating the AES or DES algorithm output and represented the malware payload as custom PNG, JPG or a sound file. Furthermore, the authors demonstrated that such payloads remain undetected with the anti-malware. Malware authors may well be interested in hiding malware \texttt{APK} inside the assets or resource folders. Besides, malware author can develop a genuine \texttt{APK} including the customized JPG as its asset. The malicious app loads the asset at runtime to execute malicious behavior. Following protection, obfuscation, and optimization techniques are interesting:
%\vspace{0.2 cm}

\begin{enumerate}
\item Using Proguard from Android SDK to protect apps proprietary logic: Proguard performs variable renaming, leaving the class name un-obfuscated. Hence, it is easy to identify the obfuscated class as its methods are modified but the class name remains un-obfuscated.
%\vspace{0.1 cm}

\item Strings encoded with Base64: 
malware authors exploit various forms of string transformations such as string encryption using arrays, non-ASCII character replacement or, hides resource files inside the strings encrypted with Base64 encoding methods. Thus, the binary data can also be hidden as Base64 encoded strings. 
%\vspace{0.1 cm}

\item Dynamic loading: 
it allows the external code to be downloaded at runtime in an APK. This technique has also been used by malware authors as discussed in Table~\ref{table:malware_obfuscation}. For the initial automation phase, its presence was only detected by pattern matching check of the classes for the packages.

%\vspace{0.1 cm}
\item Native Code: identifies the presence of native code by filtering the class definition table to verify the usage of application code accessing the system related information and resources or interfacing with the runtime environment. 
%\vspace{0.1 cm}

\item Reflection: the classes definition table was filtered for the presence of the Java reflection packages for access to methods, fields and classes.

%\vspace{0.1 cm}
\item Header size: bytecode injection inside the \texttt{classes.dex} header can be exploited by taking advantage of discrepancy between the dalvik bytecode documentation and the file. 
%\vspace{0.1 cm}

\item Encoding: it verifies the presence of mixed endianness with a flag check.
%\vspace{0.1 cm}

\item Cryptographic code: \texttt{javax.crypto} and \texttt{java.security.spec} packages provide facilities to implement encryption/decryption in classes and interface application to study  misuse of cryptographic functionality. 
\end{enumerate}

%\vspace{-0.05 mm}

\noindent Malware authors have misused the existing code protection and obfuscation methods to evade the commercial anti-malware. Table~\ref{table:malware_obfuscation} summarizes malware obfuscation chronology and illustrates the methods and tools used by to obfuscate \texttt{APK} files. Malware writers leveraged custom encryption, string encryption, and URL encryption. They encrypted network communications and encoded URLs. The recent malware apps exploit sophisticated techniques like steganography and code obfuscation tools to harden the malware reverse engineering.

\indent In 2013, malware developer used Dexguard to evade the anti-malware. The malware writers pushed Javascript payloads inside the resource folder and encrypted non-Dalvik code in the obfuscated apps. Some malware samples developed in the year 2014 employed online Packing tool services apart from string encryption and obfuscation. Additionally, malware writer encrypted \texttt{data.xml} files inside APK archive to evade anti-malware and harden reverse engineering. Dendroid~\cite{dendroid} is a stealth remote administration toolkit employing hidden behavior sending premium-rate SMS, voice calls, recording audio video without user consent. The Trojan evades the existing commercial anti-malware techniques. Elish et al.~\cite{Elish:2015:PUD:2775757.2776272} proposed user-intention based anomaly detector to detect such stealth malicious apps.

\begin{table*}
%\rowcolors{1}{white}{lightgray}
\scriptsize
\centering
\scalebox{0.81}{
\begin{tabular}{@{}llll@{}}

\toprule
\multicolumn{1}{c}{\textbf{Malware}} & \multicolumn{1}{c}{\textbf{Year}} & \multicolumn{1}{c}{\textbf{\begin{tabular}[c]{@{}c@{}}Obfuscation/Encryption/\\ Protection/Optimization Method\end{tabular}}} & \multicolumn{1}{c}{\textbf{Tool/Technique used}} \\ \midrule

SlemBunk                        & 2015                              & Code Obfuscation & Class, Method and Field Obfuscation \\ \midrule

Trojan.Dropper.RealShell                        & 2015                              & Custom APK Obfuscation                                                                                                      & Stores files in Assets Folder                                     \\ \midrule

Dendroid.A!tr                        & 2014                              & Obfuscated with Dexguard                                                                                                      & Dexguard Tool                                    \\ \midrule
SmsSend.ND                           & 2014                              & Packed with APKProtect Packer                                                                                                 & Code Packing Tool                                \\ \midrule
Freejar.B                            & 2014                              & Packed with Bangcle Packing service                                                                                           & Online code packing Service                      \\ \midrule
RuSMS.AO                             & 2014                              & Strings obfuscated, using Adobe Airpush like name                                                                             & Custom string encryption                         \\ \midrule
SmsSpy.HW!tr.spy                     & 2014                              & ``data.xml'' file is encrypted with Blowfish algorithm                                                                          & Custom symmetric encryption                      \\ \midrule
Agent.BH!tr.spy                      & 2014                              & Sends encrypted emails using TL security                                                                                      & Custom encryption                                \\ \midrule
Rmspy.A!tr                           & 2013                              & Obfuscated with Dexguard                                                                                                      & Dexguard obfuscator                              \\ \midrule
oBad                                 & 2013                              & Obfuscated with Dexguard                                                                                                      & Dexguard obfuscator                              \\ \midrule
Android.Ginmaster                    & 2013                              & Custom String encryption                                                                                                      & Custom encryption                                 \\ \midrule
Android/GinMaster.L                  & 2013                              & String obfuscation with string table in array                                                                                 & Custom encryption                                \\ \midrule
Stels.A!tr                           & 2013                              & Custom encoding URL text with Base64                                                                                          & Custom encoding                                  \\ \midrule
Pincer.A!tr.spy                      & 2013                              & Caeser cipher to hide text and telephone No.                                                                                  & Symmetric encryption algorithm                   \\ \midrule
GinMaster.B                          & 2013                              & Encrypts IMEI, IMSI and strings with Triple DES                                                                               & Custom symmetric encryption                      \\ \midrule
FakeDefend.A!tr                      & 2013                              & Encrypted fake infections with AES algorithm                                                                                  & Block cipher encryption                          \\ \midrule
FakePlay.B!tr                        & 2013                              & Non Dalvik, Javascript payload in resources                                                                                   & Non Dalvik code encryption                       \\ \midrule
SmsSend.N                            & 2012                              & Obfuscated with Proguard                                                                                                       & Proguard obfuscator                              \\ \midrule
Plankton.B!tr                        & 2012                              & Obfuscated with Proguard                                                                                                       & Proguard obfuscator                              \\ \midrule
DroidKungFu.D!tr                     & 2012                              & Obfuscated with Proguard                                                                                                       & Proguard obfuscator                              \\ \midrule
%Android/SmsZombie.A!tr               & 2012                              & Hidden Package, Jar file in resource \& assets                                                                                & Dalvik \& non Dalvik obfuscation                 \\ \midrule
FakeInst.A!tr.dial                   & 2012                              & PNG file stores SMS text body and phone numbers                                                                               & Steganography                                    \\ \midrule
%Android/Plankton                     & 2012                              & License Verification Library (LVL)  obfuscation                                                                               & Android obfuscation                              \\ \midrule
NotCompatible.Android!tr.bdr         & 2012                              & Encrypted C\&C URL in resource folder with AES                                                                                & Block Cipher encryption                          \\ \midrule
DroidKungFu.G!tr                     & 2012                              & ELF payload stored as ``mylogo.jpg ''                                                                                             & Non Dalvik code encryption                       \\ \midrule
Pjapps.A!tr                          & 2011                              & Encoded URL                                                                                                                   & URL encryption                                   \\ \midrule
Android/SmsSpy.HW!tr                 & 2011                              & Encrypted with symmetric key Blowfish algorithm                                                                               & Symmetric encryption                            \\ \midrule
Android/RootSmart                    & 2011                              & Symmetric key encryption DES, AES and Blowfish                                                                                & Block cipher encryption                          \\ \midrule
BaseBridge.A!tr                      & 2011                              & String encrypted in an array                                                                                                  & Encrypted strings                                \\ \midrule
%JSmsHider.A!tr                       & 2011                              & Variable name and string encryption                                                                                            & Custom encryption                                
Geinimi.A!tr                         & 2011                              & Encrypted network communication \& obfuscated codes                                                                           & custom encryption and obfuscation                %\\ \midrule
%Android/Geinimi                      & 2010                              & Encrypted with Data Encryption Standard                                                                                        & Symmetric encryption                             
\\ \bottomrule
\end{tabular}
}
\caption{\label{table:malware_obfuscation}Malware Obfuscation chronology~\protect\cite{blackhat:2014,vb:2014,apvrille:2013,Zhou:20122,Lookout:2013,bangcle}}
    
\end{table*}

\vspace{-0.2cm}
\section{Existing Surveys and related work}
\label{sec_relatedsurveys}
\par Shabtai et al.~\cite{10.1109/MSP.2010.2} proposed an Android threat taxonomy on Android platform. In~\cite{Vidas:2011:YDB:2028052.2028062}, the authors survey different attack vectors, and discuss the attack taxonomy on Android. In~\cite{zhou:dissecting}, authors, carried out Android malware characterization of 49 Android malware families. The authors reported simple obfuscation techniques employed by malware such as Anserver, and Bgserv. Enk et al.~\cite{Enck:2011:SAA:2028067.2028088} discussed the existing research, primarily targeting Android platform. The authors reviewed the Android platform security and app analysis methods. Furthermore, the authors discussed limitations of analysis techniques like rule-based detection, ex-filtration of sensitive information and inter-application privilege escalation attacks. However, the code obfuscation and protection techniques are not covered in the proposed analysis techniques. The proposed review gives an extensive insight into the obfuscation techniques and code protection methods. Furthermore, the review compares various code obfuscation and de-obfuscation techniques employed by malware authors and anti-malware industry.
%\vspace{0.15cm}

\par Tangil et al.~\cite{6657497} discuss the evolution of mobile malware, their infection and distribution techniques and detail them with different case studies. The authors survey the greyware and malicious app detection techniques between 2010 and 2013. Furthermore, the authors discuss various research problems, briefly discussing the impact of malware detection. In~\cite{6999911}, authors discuss the Android security issues, malware penetration and various defense methods for app analysis and mobile platform security. Furthermore, the authors briefly review obfuscation techniques. However, they concentrate more on malicious repackaging, a common problem with Android apps. 
%\vspace{0.15cm}

\par In~\cite{conf/iait/Kovacheva13}, authors focus on developing efficient Dalvik bytecode obfuscation techniques. They study the Google Play apps to identify the feasibility of availing the source code with reverse engineering tools. Authors propose efficient obfuscator design to defeat the existing de-obfuscation tools (i.e., smali, dedexer, ded). Schulz et al.~\cite{325schulz} performed Android bytecode de-obfuscation feasibility. The authors evaluated analysis methods to automate de-obfuscation of Dexguard obfuscated code. Rastogi et al.~\cite{rcj14} evaluated commercial anti-malware against trivial code obfuscation techniques. Faruki et al.~\cite{Faruki:trustcom} compare the performance of anti-malware, and static analysis tools against popular \texttt{x86} transformation attacks. Harrison et al.~\cite{rowenhollaway} investigate the effect of code obfuscation on Android. They evaluate the limitations of reverse engineering tools against app repackaging. 

\par Schrittwieser et al.~\cite{Schrittwieser:2016:PST:2911992.2886012} evaluate smartphone code protection techniques. The authors analyze and evaluate software de-obfuscation techniques. The survey is more general targeting software protection techniques and analysis methods. However, our target is, Android specific obfuscation techniques.

The proposed review is a comprehensive discussion on source code obfuscation, code protection, Android specific obfuscation, and code protection tools. To the best of our knowledge, we are the first to investigate code protection and malware obfuscation techniques for the Android platform. We discuss Collberg taxonomy~\cite{Collberg02watermarking}, and expand source code, and bytecode obfuscation taxonomy.

\section{FUTURE RESEARCH DIRECTIONS} \label{sec_frd}
\noindent Malware such as \texttt{Android/DroidCoupon.A!r}, and \texttt{AndroidSmsZombie.A!.tr} hide the malicious native payloads as JPG, or PNG files~\cite{Apvrille:2014},~\cite{vb:2014}. However, the assets are payloads just named as graphic files. Making fake payload with such tricks is prevalent on Android malware applications. Furthermore, authors in~\cite{blackhat:2014} developed a Proof-of-Concept (PoC) AngeCryption~\cite{blackhat:2014} to illustrate the possibility of encrypting any given input stored as an image (JPG, PNG). In particular, an attacker can develop a benign APK file to hide a malicious image inside resource or asset to evade the anti-malware.

\indent The unsuspected image containing malicious payload can be used to execute the malicious code. Such an attack may not be noticed at all, as the APK does not contain obfuscated, protected or wrapped content. 

\indent AngeCryption has demonstrated a PoC on the latest Android OS version. Thus, a malicious \texttt{dex} file can be embedded inside an image. Furthermore, it can be obfuscated with obfuscation tool. The dynamic code loading techniques can be used to execute the malicious payload. At present, methods to detect such attacks are not available. The functioning of such payloads cannot be determined before runtime image decryption. The suggested remedy that we aim to investigate in the future are: (i) keep tab on an APK that decrypts its resource or assets (such apps can be analyzed dynamically to identify suspicious behavior); (ii) analyze image decryption to an APK as malicious.

\indent The Android devices have constrained processing and limited storage. Obfuscation techniques does have an adverse impact on battery consumption. The power management is an important issue to identify impact of code level modifications. The Android Obfuscation has a APK statistical significance~\cite{sahin15jsep}. An important future work is to consider a large set of obfuscated APK empirical evaluation. The same can be extended to different mobile OS and devices. Since the developers do not have access to tools like CARAT~\cite{Peltonen201671}, they cannot identify the impact on energy consumption. The ability to identify the impact is important for resource constrained Android devices.

\indent The Android apps have a lot of user interaction and string usage. The malware authors use string encryption and obfuscation techniques to hide the plaintext strings. In this paper, we have discussed notable malicious apps using such encryption techniques. The authors can implement inter-component communication based inter-procedural static analysis to reconstruct the encrypted strings to obtain insight into the string information. Furthermore, authors in~\cite{MSR-2014-VasquezHBP} empirically evaluated third-party library and obfuscated code usage. To monitor the apps, we propose to identify the third-party libraries to identify APK cloning. The common use of Google advertisement network, Facebook ad libraries impacts the categorization. As a part of future work, we plan to delink the library code from APK files and evaluate obfuscated code.

\indent The existing academic code obfuscation research is heavily concentrated more towards analysis of obfuscated malicious applications~\cite{Zeng:2013:ORB:2508859.2516664}~\cite{Armknecht:2013:SFA:2508859.2516650} ~\cite{Xing:2014:UYA:2650286.2650760}~\cite{180374}. The relevant literature evaluates obfuscation techniques prominently among malicious applications. The real identification of obfuscated code among the normal programs which is important for software protection, is ignored. The non-malicious code reverse engineering is largely unexplored. Targeting program obfuscation and related techniques for protecting the digital rights is an interesting future direction. Inspite of the existing research on obfuscation, evaluation matrices to verify the existing obfuscation technique resilience are not available. Formal analysis techniques to evaluate obfuscation and de-obfuscation techniques is still not available. Hence, we summarize code obfuscation, de-obfuscation tools and techniques to understand the effect in isolation. It would be interesting to combine different class of obfuscation techniques, and evaluate existing de-obfuscation tools. 

\vspace{0.2cm}
\section{CONCLUSION} \label{sec:con}
\indent Android, currently the most popular mobile OS platform is eight-year-old. The growth and commercial value has attracted the research community and malware authors alike. Since the mobile OS is fast evolving, code protection techniques are implemented by the app developers to harden the reverse engineering of the code propriety. On the other hand, malware authors are using the protection techniques to delay the code reverse engineering. 
\par In this survey, we address important and specific questions about obfuscation and code protection techniques on mobile platform. In the existing obfuscation research, evaluation techniques to assess resilience of obfuscation are still not available. Code analysis and de-obfuscation techniques have similar limitations. We performed review of the existing code obfuscation and analysis techniques isolated from one another. We discuss the details of code protection, optimization and obfuscation technique. Furthermore, we explore custom code protection techniques employed by malware authors to hide malicious payloads. Obfuscation tools and techniques also depends on availability of resources for reverse engineering. Existing tools (e.g., Androguard, JEB, dex2jar) de-obfuscate custom code examples; however, they fail to decode real-world programs. The complexity of a problem may outdo the existing resources. Hence, simple obfuscation techniques can be effective on resource constrained devices. This is one of the reason of its popularity among malware authors. The ongoing challenge between code protection and analysis techniques is growing. Specific obfuscation methods are effective in certain situations. Given time and effort, existing obfuscation techniques can be decoded by human analyst. 

\newpage
\appendix
\section{Obfuscation code examples}
In the following, we discuss \texttt{FakeInstaller}, stealth Android malware employing different class of obfuscation to evade anti-malware. Listing~\ref{confida_motivation} reverse engineered code of \texttt{FakeInstaller}~\cite{fakeinstaller,TUD-CS-2015-0031,jeb} at line number 1 checks for the presence of emulator, an alibi for development environment, or automated analysis system. In line number 9, class and method names are obfuscated to erase the program semantics. For example, a random string value ``VQIf3AInVTTnSaQI+R]KR9aR9'', is decrypted to Android \texttt{android.telephony.SmsManager} class. This class is loaded using reflection API at runtime to evade static analysis. The class sends premium-rate SMS without explicit user consent. The string ``BaRIta*9caBBV]a'' is decrypted to \texttt{SendTextMessage} method. Furthermore, in line number 21 \texttt{getMethod} sends the SMS using the text from the parameters p1 and p2, declared in line number 1. Here, the use of multiple obfuscation, evasion and code protection techniques successfully evades the static analysis.

%\begin{minipage}{.48\textwidth}
\definecolor{mygreen}{rgb}{0,0.6,0}
\lstset{
    captionpos=b,
    caption={Fakeinstall Obfuscation~\protect\cite{fakeinstaller,TUD-CS-2015-0031}},
    label=confida_motivation,
    keywordstyle=\color{blue},
    commentstyle=\color{mygreen},
    basicstyle=\fontsize{7}{8}\selectfont\ttfamily,
    frame=single,
    numbers=left,
    belowcaptionskip=2\baselineskip,
    tabsize=2,
}
\begin{lstlisting}[language=Java,xleftmargin=.05\textwidth, xrightmargin=.05\textwidth]
// Fakeinstall obfuscated code
public static boolean gdadfjrj(String p1,String p2){ [...]

// Anti analysis check to evade emulator

if (zhfdghfdgd()) return;

// Get class instance
Class clz = Class.

forName(gdadfjrj.gdafbj("VQIf3AInVTTnSaQI+R]KR9aR9"));
Object localObject = clz.getMethod(gdadfjrj.gdadfjrj("]a9maFVM.9"), newClass[0])
.invoke(null, new Object[0]);

// Get the method name
String s = gdadfjrj.gdadfjrj("BaRIta*9caBBV]a");

// Build parameter list
Class c = Class.forName(gdadfjrj.gdadfjrj("VQIf3AInVTTnSaQI+R]KR9aR9"));
Class[] arr = new Class[] 
{nglpsq.cbhgc, nglpsq.cbhgc, nglpsq.cbhgc, c, c };

// Reflection for invoking the method to send SMS
clz.getMethod(s, arr).invoke(localObject, newObject[] { p1, null, p2, null,null });
\end{lstlisting}
%\end{minipage}\hfill

\vspace{-0.5cm}
The partial code snippet in Listing~\ref{zitmo_malware} is a variant of zitmo~\cite{Li2015}. As illustrated, when the SMS is received, framework callback \texttt{onReceive()} is invoked to stop the broadcast to default SMS app. The \texttt{abortBroadcast()} method aborts the current broadcast. Then, an intent that carries the SMS is launched inside the MainService, a background service task. Further, the stored SMS from the intent is accumulated in the array called ``pdus''. The sender identification and message parts are extracted by \texttt{getOriginatingAddress()} and \texttt{getMessageBody()} methods. Furthermore, the available values of the ``pdus'' object are stored inside variables s1, s2 along with the device Id using method \texttt{getDeviceId()}. The information is encoded inside the \texttt{UrlEncodedFormEntity} object. Further, the constant URL string is then encoded with \texttt{setEntity()} to post the data using \texttt{execute()} method. \newpage

%\begin{minipage}{.48\textwidth}
\definecolor{mygreen}{rgb}{0,0.6,0}
\lstset{
    captionpos=b,
    caption={SMS Obfuscation \& IMEI exfiltration~\protect\cite{Li2015}},
    label=zitmo_malware,
    keywordstyle=\color{blue},
    commentstyle=\color{mygreen},
    basicstyle=\fontsize{6.5}{7.5}\selectfont\ttfamily,
    frame=single,
    numbers=left,
    belowcaptionskip=2\baselineskip,
    tabsize=2,
}
\begin{lstlisting}[language=Java,xleftmargin=.05\textwidth, xrightmargin=.05\textwidth]
public class mysmsReceiver extends BroadcastReceiver
{
 public void onReceive (context pcontext, intent pintent)
 {
		bundle localBundle=pintent.getExtras();
       if((localBundle != null)&&(localBundle.containsKey("pdus")))
	{ 
		abortBroadcast();
		Intent.targetService=newIntent(pcontex, MainService.class);
    		targetService.putExtra("pdus",localBundle);
    		pcontex.startService(ts);   
	}
 }
}

public class MainService extends Service
{
	 public int onStartCommand(Intent pintent, int pintent1, int pintent2){	
	 Bundle localBundle=pintent.getBundle("pdus");

	 SmsMessage localSMS=SmsMessage.createFromPdu("pdus");
	 String S1 = localSMS.getOriginatingAddress();
	 String S2 = localSMS.getMessageBody();
	 String S3 = localTM.getDeviceId();}

	 public void postRequest(urlEncodedFormEntity UEFE)
	{
	   String address="http://stringthrifty.com/security.jsp";
	    ...
	    DefaultHttpClient().execute(localHP, BRH);
	}
}
\end{lstlisting}
%\end{minipage}
%\end{appendix}

\vspace{-0.5cm}

% Acknowledgments
\begin{acks}
The work of Parvez Faruki, Manoj Singh Gaur and Vijay Laxmi is supported by department of Information Technology, Government of India project grant ``SAFAL-Security Analysis Framework for Android pLatform''. Hossein Fereidooni is supported by Deutsche Akademische Austauschdienst (DAAD fellowship). Mauro Conti is supported by Marie Curie fellowship funded by the European Commission under the agreement No. PCIG-GA-2012-321980. This work is partially supported by the TENACE PRIN Project 20123P34XC funded by Italian MIUR, and ``Tackling Mobile Malware with Innovative Machine Learning Techniques'' funded by University of Padua.
\end{acks}

\bibliographystyle{ACM-Reference-Format-Journals}
\bibliography{sigproc.bib}

%\end{document}

% History dates
%\received{February 2007}{March 2009}{June 2009}

% Electronic Appendix
%\elecappendix

%\medskip

\end{document}